\documentclass{article}

    \PassOptionsToPackage{authoryear, round}{natbib}

    \usepackage[preprint]{neurips_2023}

\usepackage{setspace}
\usepackage[utf8]{inputenc} 
\usepackage[T1]{fontenc}    

\usepackage[hyphens]{url}
\usepackage{hyperref}
\usepackage[hyphenbreaks]{breakurl}

\usepackage{url}            
\usepackage{booktabs}       
\usepackage{amsfonts}       
\usepackage{nicefrac}       
\usepackage{microtype}      
\usepackage{xcolor}         

\usepackage{tcolorbox}
\usepackage{enumitem}

\usepackage{footmisc} 

\usepackage{soul}
\usepackage{longtable}
\usepackage{cleveref}

\usepackage{array} 
\providecommand{\tightlist}{%
  \setlength{\itemsep}{0pt}\setlength{\parskip}{0pt}}

\hypersetup{
    colorlinks,
    linkcolor={blue!80!black},
    citecolor={blue!80!black},
    urlcolor={blue!80!black}
}

\definecolor{customlightblue}{HTML}{ddebff} 
\definecolor{customdarkblue}{HTML}{BED8FF} 

\definecolor{customlightgreen}{HTML}{dfffdb}
\definecolor{customdarkgreen}{HTML}{c4f2bd}

\setuldepth{A} 

\newlength{\nameaffilspace}
\newcommand{\boxspace}{-0.15cm}

\title{Taking AI Welfare Seriously}

\author{
Robert Long$^*$ \\[\nameaffilspace]
Eleos AI\\
\And
Jeff Sebo$^*$\\[\nameaffilspace]
New York University \\
\AND
Patrick Butlin$^\dagger$ \\[\nameaffilspace]
University of Oxford \\
\And
Kathleen Finlinson$^\dagger$ \\[\nameaffilspace]
Eleos AI \\
\And
Kyle Fish$^\dagger$$^\mathsection$ \\[\nameaffilspace]
Eleos AI, Anthropic \\
\AND
Jacqueline Harding$^\dagger$ \\[\nameaffilspace]
Stanford University \\
\And
Jacob Pfau$^\dagger$ \\[\nameaffilspace]
New York University \\
\And
Toni Sims$^\dagger$ \\[\nameaffilspace]
New York University \\
\AND
Jonathan Birch$^\ddagger$ \\[\nameaffilspace]
London School of Economics \\
\And
David Chalmers$^\ddagger$ \\[\nameaffilspace]
New York University \\
}

\begin{document}

\maketitle

\renewcommand{\thefootnote}{$*$}
\footnotetext{Lead and corresponding authors. \{\url{robert@eleosai.org}, \url{jeffsebo@nyu.edu}\}}
\renewcommand{\thefootnote}{$\dagger$}
\footnotetext{Main authors.}
\renewcommand{\thefootnote}{$\ddagger$}
\footnotetext{Contributing authors.}
\renewcommand{\thefootnote}{$\mathsection$}
\footnotetext{Work performed while at Eleos AI, prior to joining Anthropic in Fall 2024.}

\begin{abstract}
    In this report, we argue that there is a realistic possibility that some AI systems will be conscious and/or robustly agentic in the near future. That means that the prospect of AI welfare and moral patienthood — of AI systems with their own interests and moral significance — is no longer an issue only for sci-fi or the distant future. It is an issue for the near future, and AI companies and other actors have a responsibility to start taking it seriously. We also recommend three early steps that AI companies and other actors can take: They can (1) acknowledge that AI welfare is an important and difficult issue (and ensure that language model outputs do the same), (2) start assessing AI systems for evidence of consciousness and robust agency, and (3) prepare policies and procedures for treating AI systems with an appropriate level of moral concern. To be clear, our argument in this report is not that AI systems definitely are — or will be — conscious, robustly agentic, or otherwise morally significant. Instead, our argument is that there is substantial uncertainty about these possibilities, and so we need to improve our understanding of AI welfare and our ability to make wise decisions about this issue. Otherwise there is a significant risk that we will mishandle decisions about AI welfare, mistakenly harming AI systems that matter morally and/or mistakenly caring for AI systems that do not.
\end{abstract}

\renewcommand{\thefootnote}{\arabic{footnote}} 

\newpage 
\setcounter{tocdepth}{3}
\tableofcontents

\setcounter{footnote}{0}

\newpage 

\vspace{-0.5cm}

\hypertarget{introduction}{%
\section{Introduction}\label{introduction}}

\vspace{-0.5cm}

\hypertarget{a-transitional-moment-for-ai-welfare}{%
\subsection{A transitional moment for AI
welfare}\label{a-transitional-moment-for-ai-welfare}}

\vspace{-0.3cm}

In this report, we argue that there is a realistic possibility that some AI systems will be conscious and/or robustly agentic, and thus morally significant, in the near future.\footnote{By “near term” or “near future” we mean roughly within the next decade, so by around 2035, but nothing in our argument or recommendations depends on this exact timeline.
} We also argue that AI companies have a responsibility
to acknowledge that AI welfare\footnote{As we discuss below, by ‘AI welfare’ we mean AI systems with morally significant interests and, relatedly, the capacity to be benefited or harmed.} is a serious issue; start assessing their AI
systems for welfare-relevant features; and prepare policies and
procedures for interacting with potentially morally significant AI
systems. Plausible
philosophical and scientific theories, which accord with mainstream
expert views in the relevant fields, have striking implications for this
issue, for which we are not adequately prepared. We need to take steps
toward improving our understanding of AI welfare and making wise
decisions moving forward.

\vspace{-0.1cm}

We release this report during a transitional moment for AI welfare. For
most of the past decade, AI companies appeared to mostly treat AI
welfare as either an imaginary problem or, at best, as a problem only
for the far future. As a result, there appeared to be little or no
acknowledgment that AI welfare is an important and difficult issue;
little or no effort to understand the science and philosophy of AI
welfare; little or no effort to develop policies and procedures for
mitigating welfare risks for AI systems if and when the time comes;
little or no effort to navigate a social and political context in which
many people have mixed views about AI welfare; and little or no effort
to seek input from experts or the general public on any of these
issues.

\vspace{-0.1cm}

Recently, however, some AI companies have started to acknowledge that AI
welfare might emerge soon, and thus merits consideration today. For
example, Sam Bowman, an AI safety research lead at Anthropic, recently argued (in a personal capacity) that Anthropic needs to “lay the groundwork for AI welfare commitments,” and to begin to “build out a defensible initial understanding of our situation, implement low-hanging-fruit interventions that seem robustly good, and cautiously try out
formal policies to protect any interests that warrant protecting.”\footnote{\cite{bowman_checklist_2024}} Google recently
announced that they are seeking a research scientist\footnote{\cite{google_job_ad}} to work on ``cutting-edge societal questions around machine
cognition, consciousness and multi-agent systems''. High-ranking members
of other companies have expressed concerns as
well.\footnote{See \cite{long_experts_2024} for more examples.}

\vspace{-0.1cm}

This growing recognition at AI companies that AI welfare is a credible
and legitimate issue reflects a similar transitional moment taking place
in the research community. Many experts now believe that AI welfare and
moral significance is not only possible in principle, but also a
realistic possibility in the near future.\footnote{For work on AI
  welfare as a near-term issue, see 
\cite{birch_edge_2024,schwitzgebel_coming_2023,schwitzgebel_full_2023,chalmers_could_2023,sebo_moral_2023,goldstein_case_2024,bradley_ai_2024,dung_saving_nodate}.
  } And
even researchers who are skeptical of AI welfare and moral significance
in the near term advocate for caution; for example, leading
neuroscientist and consciousness researcher Anil Seth writes, ``While
some researchers suggest that conscious AI is close at hand, others,
including me, believe it remains far away and might not be possible at
all. \textbf{But even if unlikely, it is unwise to dismiss the
possibility altogether} [emphasis ours].''\footnote{\cite{seth_why_2023}.This accords with a
  number of papers and thinkers which discuss the high stakes of
  underattributing moral status and how to deal with moral status given
  uncertainty, including \cite{chan_ethical_2011,birch_animal_2017,sebo_moral_2018,dung_tests_2023,ladak_what_2024,goldstein_case_2024}.
  }

\vspace{-0.15cm}

Our aim in this report is to provide context and guidance for this
transitional moment.\footnote{This report is the first output of a broader research project. In future work, we
  will release a research agenda about how
  AI companies and others can assess AI systems for
  consciousness and robust agency, and 
  develop
  policies and procedures for treating AI systems with an appropriate
  level of moral concern. Given the early stage of this field, this
  report may also be a living document that is updated periodically.} To improve our understanding and decision-making
regarding AI welfare, we need more precise empirical frameworks for
evaluating AI systems for consciousness, robust agency, and other
welfare-relevant features. We also need more precise normative
frameworks for interacting with potentially morally significant AI
systems and for navigating disagreement and uncertainty about these
issues as a society.\footnote{According to one survey of public opinion
  (\cite{colombatto_folk_2024}, the majority of the public is already
  willing to attribute some chance of consciousness to large language
  models. Experts have a responsibility not only to research AI welfare
  but to disseminate that research publicly.} This report outlines
several steps that AI companies can take today in order to start
preparing for the possible emergence of morally significant AI systems
in the near future, as a precautionary measure.\footnote{The
  ``precautionary principle'' is a term of art for a particular view
  about decision-making under uncertainty, (see \cref{the-risks-of-mishandling-ai-welfare}). But here we mean ``precautionary''
  in the ordinary sense of the word.}

\vspace{-0.20cm}

We begin in \cref{introduction} by explaining why AI welfare is an important and
difficult issue. Leaders in this space have a responsibility to
understand this issue as best they can, because errors in either
direction --- either over-attributing or under-attributing moral
significance to AI systems --- could lead to grave harm. However,
understanding this issue will be challenging, since forecasting the
mental capacities and moral significance of near-future AI systems
requires improving our understanding of topics like the nature of
consciousness, the nature of morality, and the future of AI. It also
requires overcoming well-known human biases, including a tendency to
both over-attribute and under-attribute capacities like consciousness to
nonhuman minds.

\vspace{-0.25cm}

In \cref{routes-to-near-term-ai-welfare}, we argue that given the best information and arguments
currently available, there is a realistic possibility of morally
significant AI in the near future. We focus on two mental capacities
that plausibly suffice for moral significance: consciousness and robust
agency. In each case, we argue that caution and humility require
allowing for a \textbf{realistic possibility} that (1) this capacity
suffices for moral significance \emph{and} (2) there are certain
computations that (2a) suffice for this capacity \emph{and} (2b) will
exist in near-future AI systems. Thus, while there might not be
certainty about these issues in either direction, there is a \emph{risk}
of morally significant AI in the near future, and AI companies have a
responsibility to take this risk seriously now.\footnote{To be more
  precise, the risk is that morally significant AI will be created
  \emph{and harmed or wronged}.}

\vspace{-0.1cm}
\begin{tcolorbox}[
  width=\textwidth,
  colback=customlightblue,
  colbacktitle=customdarkblue,
  sharp corners,
  boxrule=1pt, 
  title={\vspace{0.1cm}We argue that, according to the best evidence currently available, there is a realistic possibility that some AI systems will be welfare subjects and moral patients in the near future.\vspace{0.1cm}},
  coltitle=black,
  titlerule=0pt,
  left=1mm,
  right=1mm,
]
{
\setstretch{1.0}
\scalebox{.95}[1]{
\textbf{\ul{Consciousness route to moral patienthood.}} There is a \textbf{realistic, non-negligible possibility} that:}
\begin{enumerate}
    \tightlist 
    \item \textbf{Normative}: Consciousness suffices for moral patienthood,
    \emph{and}
    \item \textbf{Descriptive}: There are computational features --- like a
    global workspace, higher-order representations, or an attention schema
    --- that \emph{both}:
    \begin{itemize} 
        \tightlist 
        \item[a.] Suffice for consciousness, \emph{and}
        \item[b.] Will exist in some near-future AI systems. \\
    \end{itemize}
\end{enumerate}
\vspace{-0.3cm}
\scalebox{.95}[1]{
\textbf{\ul{Robust agency route to moral patienthood.}} There is a \textbf{realistic, non-negligible possibility} that:}
\begin{enumerate}
    \tightlist 
    \item \textbf{Normative}: Robust agency suffices for moral patienthood,
    \emph{and}
    \item \textbf{Descriptive}: There are computational features --- like
    certain forms of planning, reasoning, or action-selection --- that
    \emph{both}:
    \begin{itemize}
        \tightlist 
        \item[a.] Suffice for robust agency, \emph{and}
        \item[b.] Will exist in some near-future AI systems.
    \end{itemize}
\end{enumerate}
}
\end{tcolorbox}
\vspace{\boxspace}

We close, in \cref{recommendations-for-ai-companies}, by presenting three procedural steps that AI
companies can take today, in order to start taking AI welfare risks
seriously. Specifically, AI companies can (1)
\textbf{acknowledge} that AI welfare is an issue, (2) take steps to
\textbf{assess} AI systems for indicators of consciousness, robust
agency, and other potentially morally significant capacities, and (3)
take steps to \textbf{prepare} policies and procedures that will allow
them to treat AI systems with an appropriate level of moral concern in
the future. In each case we also present principles and potential
templates for doing this work, emphasizing the importance of developing
ecumenical, pluralistic decision procedures that draw from expert and
public input.

\begin{tcolorbox}[
  width=\textwidth,
  colback=customlightgreen,
  colbacktitle=customdarkgreen,
  sharp corners,
  boxrule=1pt, 
  title={\vspace{0.1cm}\textbf{Recommendations.} We recommend that AI companies take these minimal first steps towards taking AI welfare seriously.\vspace{0.1cm}},
  coltitle=black,
  titlerule=0pt,
  left=1mm,
  right=1mm,
]
\textbf{\ul{Acknowledge.}} Acknowledge that AI welfare is an important and
  difficult issue, and that there is a realistic, non-negligible chance
  that some AI systems will be welfare subjects and moral patients in
  the near future. That means taking AI welfare seriously in any
  relevant internal or external statements you might make. It means
  ensuring that language model outputs take the issue seriously as well.\\
  
\textbf{\ul{Assess.}} Develop a framework for estimating the probability
  that particular AI systems are welfare subjects and moral patients,
  and that particular policies are good or bad for them. We have
  templates that we can use as sources of inspiration, including the
  ``marker method'' that we use to make estimates about nonhuman
  animals. We can consider these templates when developing a
  probabilistic, pluralistic method for assessing AI systems.\\
  
\textbf{\ul{Prepare.}} Develop policies and procedures that will allow AI
  companies to treat potentially morally significant AI systems with an
  appropriate level of moral concern. We have many templates to
  consider, including AI safety frameworks, research ethics frameworks,
  and forums for expert and public input in policy decisions. These
  frameworks can be sources of inspiration --- and, in some cases, of
  cautionary tales.\\

  \emph{These steps are necessary but far from sufficient. AI companies and other actors\footnotemark\ have a responsibility to start considering and mitigating AI welfare risks.}

\end{tcolorbox}

\footnotetext{In our recommendations, we sometimes use a collective "we". In those moments, we are referring to the constellation of actors that have a role to play in this work, including researchers, companies, and governments.\label{ftn:rec-caveat}}

Before we begin, it will help to emphasize five important features of
our discussion. First, our discussion will concern whether near-future
AI systems might be \emph{welfare subjects} and \emph{moral patients}.
An entity is a \textbf{moral patient} when that entity \emph{morally
matters for its own sake},\footnote{See \cite{kamm_intricate_2007} for an influential
  definition of moral patienthood. ``Moral status,'' ``moral standing,''
  or ``moral considerability'' are often used interchangeably or in
  closely related ways. For more on these issues, see \cite{korsgaard_two_1983,jamieson_ethics_2008,jaworska_grounds_2021}.
  } and an entity is
a \textbf{welfare subject} when that entity has morally significant \emph{interests} and,
relatedly, is capable of being \emph{benefited} (made \emph{better off})
and \emph{harmed} (made \emph{worse off}). Being a welfare subject makes
you a moral patient --- when an entity can be harmed, we have a
responsibility to (at least) avoid harming that entity unnecessarily.
But there may be other ways of being a moral patient; our approach is
compatible with many different perspectives on these issues.

Second, our discussion often focuses on large language models (LLMs) as
a central case study for the sake of simplicity and specificity, and
because we expect that LLMs --- as well as broader systems that include
LLMs, such as language agents --- will continue to be a focal point in
public debates regarding AI welfare. But while some of our
recommendations are specific to such systems (primarily, our
recommendations regarding how AI companies should train these systems to
discuss their own potential moral significance), our three general
procedural recommendations (acknowledge, assess, and prepare) apply for
any AI system whose architecture is complex enough to at least
potentially have features associated with consciousness or robust
agency.

Third, our discussion often focuses on initial steps that AI companies
can take to address these issues. These recommendations are incomplete
in two key respects. First, AI companies are not the only actors with a
responsibility to take AI welfare seriously. Many other actors have this
responsibility too, including researchers, policymakers, and the general
public.\footnote{In this respect, AI welfare is like other high-stakes
  issues about AI development and deployment: handling AI welfare should
  not remain solely the prerogative of private corporations.} Second,
these steps are not the only steps that AI companies have a
responsibility to take. They are the \textbf{minimum necessary first
steps} for taking this issue seriously. Still, we emphasize these steps
in this report because by taking them now, AI companies can help lay the
groundwork for further steps --- at AI companies and elsewhere --- that
might be sufficient.

Fourth, our aim in what follows is not to argue that AI systems will
\emph{definitely} be welfare subjects or moral patients in the near
future. Instead, our aim is to argue that given current evidence, there
is a \emph{realistic possibility} that AI systems will have these
properties in the near future.\footnote{See \cite{goldstein_ai_nodate}: ``While \textbf{we do not claim to demonstrate
  conclusively} that AI systems have wellbeing, we argue that there is a
  \textbf{significant probability} that some AI systems have or will
  soon have wellbeing, and that this should lead us to reassess our
  relationship with the intelligent systems we create [emphasis ours]''.}
Thus, our analysis is not an expression of anything like consensus or
certainty about these issues. On the contrary, it is an expression of
\textbf{caution} \textbf{and} \textbf{humility} in the face of what we
can expect will be substantial ongoing disagreement and uncertainty.\footnote{For related work on the value of humility in AI ethics, see \cite{gellers_ai_2024}.} In
our view, this kind of caution and humility is the only stance that one
can responsibly take about this issue at this stage. It is also all that
we need to support our conclusions and recommendations
here.\footnote{Some of the authors of this report believe that near-term
  AI welfare is quite likely, and that additional measures are warranted
  at this stage. But we all believe that near-term AI welfare is, at
  minimum, likely enough to warrant the measures recommended here, and
  our aim here is to focus on arguments and recommendations about which
  we can build consensus despite our different beliefs and values.}

Finally, and relatedly, our aim in what follows is not to argue for any
particular view about how humans should interact with AI systems in the
event that they \emph{do} become welfare subjects and moral patients. We
would need to examine many further issues to make progress on this
topic, including: how much AI systems matter, what counts as good or bad
for them, what humans and AI systems owe each other, and how AI welfare
interacts with AI safety and other important issues. These issues are
all important and difficult as well, and we intend to examine them in
upcoming work. However, we do not take a stand on any of these issues in
this report, nor does one \emph{need} to take a stand on any of them to
accept our conclusions or recommendations here.\footnote{For arguments
  concerning AI moral status that use somewhat alternative
  methodological approaches than ours, see \cite{gellers_rights_2021,gunkel_machine_2012}.
  }

\hypertarget{the-risks-of-mishandling-ai-welfare}{%
\subsection{The risks of mishandling AI
welfare}\label{the-risks-of-mishandling-ai-welfare}}

When assessing the welfare and moral patienthood of nonhumans, including
other animals and AI systems, we face two kinds of risk: the risk of
\emph{over-attributing} welfare and moral patienthood to nonhumans, and
the risk of \emph{under-attributing} these properties to
nonhumans.\footnote{Arguments that uncertainty about moral status is
  dangerous because of risks of both under- and over-attribution can be
  found in, among others, \cite{christiano_when_2018,schwitzgebel_defense_2015,schwitzgebel_designing_2020,birch_edge_2024,sebo_moral_2023,dung_how_2023,shevlin_how_2021}.} Over-attribution of welfare and moral patienthood is
a false positive: mistakenly seeing, or treating, an object as a
subject, or a non-moral patient as a moral patient. Under-attribution of
these properties is a false negative: mistakenly seeing, or treating, a
subject as an object, or a moral patient as a
non-moral-patient.\footnote{\cite{de_waal_anthropomorphism_1999} similarly wrote about
  the risks of over- or under-attributing human characteristics,
  including moral status, to nonhuman animals.} Both of these mistakes
can lead to significant costs or harms in this context, and we will need
to navigate both of them with caution.\footnote{In addition to these two
  errors, we can also be mistaken about a variety of related questions
  about an entity, even assuming that they are a welfare subject and
  moral patient: how much they matter, what is good or bad for them, and
  what we owe them. These errors can also carry grave risks, and we
  discuss them further in upcoming work; see also \cite{sebo_moral_2025}. For now,
  we focus on over-attribution and under-attribution of welfare and
  moral patienthood.}

When there is a clear asymmetry between competing risks --- for example,
when false positives are far more severe than false negatives, or vice
versa --- then we might be able to mitigate risk by simply ``erring on
the side of caution'' in cases where a more complex risk assessment is
either intractable or unnecessary. But when there is at least a rough
symmetry between competing risks --- for example, when false positives
and false negatives are comparably severe --- a simple precautionary
strategy may not be possible. We may have to engage in more complex risk
assessment to the extent possible, attempting to mitigate both kinds of
risks in a reasonable, proportionate manner.

How should we think about risks involving nonhuman welfare and moral
patienthood in this context? In the case of nonhuman animals, it seems
plausible that the harms of under-attribution of welfare and moral
patienthood are often far worse than the risk of over-attribution, which
makes precautionary reasoning appropriate in those contexts. However, in
the case of AI, both errors could cause grave harm, either to humans
(and other animals) or to AI systems. Both kinds of harm could also
scale rapidly depending on the trajectory of AI development and
deployment from here. This predicament makes it difficult to simply
``err on the side of caution,'' which underscores the urgency of
improving our understanding of these issues.

On the one hand, the harm of under-attributing welfare and moral
patienthood to AI systems could be significant. When we mistakenly see a
subject as an object, we risk harming or neglecting them unnecessarily.
For example, factory farming, animal research, and other such industries
kill hundreds of billions of vertebrates and trillions of invertebrates
every year. And as evidence that these animals are welfare subjects and
moral patients has accumulated, our species has been slow to accept it,
in part because of our increasing dependence on these industries. Now
that our species is finally starting to accept this evidence, it will
take us decades to transform these industries, during which many more
animals will suffer and die unnecessarily.

In the future, similar harms could follow from under-attributing welfare
and moral patienthood to AI systems. The AI industry is currently at an
early stage of development, and depending on the path that it takes from
here, we could use even more AI systems than animals in the future, and
we could scale up our use of them even more rapidly. This is
particularly true in the current paradigm, which requires an enormous
amount of compute for training and much less for inference.\footnote{\cite{davidson_continuous_2023}} If an AI system in such a paradigm could be a welfare subject
and moral patient, then many model instances could be run after
training. Unlike with animals, the scale of the problem could increase
by orders of magnitudes more or less instantaneously.\footnote{\cite{akova_artificially_2023,bostrom_superintelligence_2014,dung_how_2023,gloor_altruists_2016,metzinger_artificial_2021,tomasik_risks_2011}}

On the other hand, the harm of over-attributing welfare and moral
patienthood to AI systems could be significant as well. First of all,
there could be substantial opportunity costs associated with this error.
At present, we lack the ability to fully care for the eight billion
humans alive at any given time, to say nothing of the quintillions of
other animals alive at any given time. If we treated an even larger
number of AI systems as welfare subjects and moral patients, then we
could end up diverting essential resources away from vulnerable humans
and other animals who really needed them, reducing our own ability to
survive and flourish. And if these AI systems were in fact merely
objects, then this sacrifice would be particularly pointless and
tragic.\footnote{\cite{wilks_robots_2010,birhane_robot_2020}}

The over-attribution of welfare and moral patienthood to AI systems
could also be actively harmful. For example, if we treated AI systems as
welfare subjects and moral patients with many of the same interests as
typical adult humans, then we could end up extending them many of the
same legal and political rights as typical adult humans, including the
right to legal and political representation and participation. This
could, in turn, empower AI systems to act contrary to our own interests,
with devastating consequences for our species\footnote{See, among
  others, \cite{bradley_ai_2024,clarke_sharing_2021,carlsmith_existential_2023},
  who notes that these risks make ``building new, very
  powerful agents who might be moral patients\ldots both a morally and
  prudentially dangerous game."} (although some have argued that
\emph{neglect} for AI systems would carry a similar risk).\footnote{\cite{salib_ai_2024,sebo_moral_2025} argue that extending legal rights
  to AI systems would help, not hinder, AI safety. We believe that this
  issue is crucial for assessing the kinds of risks discussed in this
  section, and we hope to see further research that assesses and
  compares these risks.} As with the risk of opportunity costs, this
risk would apply even if these AI systems are in fact subjects. But if
they were in fact merely objects, then accepting this risk would
likewise be particularly pointless and tragic.

By default, we should not expect our ``common sense'' intuitions about
AI welfare and moral patienthood to be reliable; we will not handle this
issue well simply by reacting to situations as they arise. We have
dispositions that can lead to under- \emph{and} over-attribution of
these properties in nonhumans, depending on the nature of the nonhumans
and our interactions with them. These include dispositions toward
\textbf{anthropomorphism}, that is, a tendency to see nonhumans as
\emph{having} human traits that they \emph{lack}. They also include
dispositions towards \textbf{anthropodenial}, that is, a tendency to see
nonhumans as \emph{lacking} human traits that they \emph{have}. Both
tendencies have caused errors regarding animals, and they will likely
have a similar effect regarding AI systems.\footnote{See \citeauthor{andrews_animal_2014}'s (\citeyear{andrews_animal_2014}) \emph{The Animal Mind} for discussion of these issues in the
  animal context.}

A number of factors make us more likely to anthropomorphize nonhumans
and, perhaps falsely, attribute consciousness and other such capacities
to them. For instance, studies suggest that we are more likely to
attribute consciousness and other such capacities to beings who move at
a similar speed as humans, rather than faster or slower.\footnote{\cite{chalmers_conscious_1996}, ch.~7; \cite{morewedge_timescale_2007}}
  We are more likely to attribute
agency to beings who have the appearance of eyes,\footnote{See \cite{fernandez-duque_is_2005}. Even infants are evidently more
  likely to treat objects as having mental states if those objects have
  eyes. See \cite{johnson_inferring_2001}.} who have distinctive motion
trajectories, and who engage in contingent interaction --- that is,
behavior that is apparently self-directed.\footnote{\cite{arico_folk_2011}.
  We note that contingent interaction is plausibly a reasonable
  criterion.} Evidence also suggests that features such as ``cuteness''
can encourage attributions of mental states and moral
patienthood.\footnote{\cite{pearce2022impact}. See \cite{campbell_ai_2024} for a brief
  popular overview of risks from unreliable intuitions about AI
  mentality.}

Many robots or chatbots are designed to appear conscious and
charismatic\footnote{There have already been examples of \emph{directly}
  optimizing chatbots to maximize user engagement: \cite{irvine_rewarding_2023}.},
and in the future, many AI systems will have bodies, life-like motion,
and (at least apparently) contingent interactions. Furthermore, unlike
nonhuman animals, AI systems are already increasingly able to hold
extremely realistic conversations, making seemingly thoughtful
contributions in realistic timeframes.\footnote{\cite{lin_duplex_2022}. As \cite{lazar_can_2024} notes, recent advances enable the creation of systems
  that ``can now offer vastly more companionable, engaging, and
  convincing simulations of friendship than has ever before been
  feasible.''} These traits do not guarantee that humans will see and
treat these systems as welfare subjects and moral patients, but they
will increase the probability of such reactions. In fact, there have
already been cases --- some prominent\footnote{For instance, AI engineer
  Blake Lemoine caused a stir by claiming that Google's AI chatbot was
  sentient in 2022. See \cite{tiku_google_2022}.} and others less so\footnote{The
  possibility and implications of AI consciousness is a popular
  discussion topic on the internet forum Reddit. See, for example, \url{https://www.reddit.com/r/ArtificialSentience/}.}
--- of humans becoming convinced that current chatbots are welfare
subjects and moral patients.

At the same time, a number of factors make us more likely to engage in
anthropodenial as well. For instance, when we consider the mechanisms
that produce nonhuman behavior --- taking what Daniel Dennett has called
taking a ``mechanistic stance''\footnote{\cite{dennett_mechanism_1973}} towards
nonhumans --- we become less likely to attribute mental states to those
nonhumans.\footnote{\cite{sims_thought_2013,nahmias_free_2007}} There
appear to be motivational factors that encourage anthropodenial as well.
For instance, those who are invested in social, political, or economic
systems that subjugate nonhumans may be more likely to view these
nonhumans as ``lesser than''. Similarly, those who find it useful to
treat nonhumans as objects may be more likely to deny that these
nonhumans are welfare subjects and moral patients.\footnote{There is
  evidence of this effect in the history of our treatment of animals.
  Often, people who eat meat are not inclined to view animals as moral
  patients. However, when they stop eating meat (even for non-moral
  reasons), they become more likely to see animals as moral patients.
  See \cite{loughnan_role_2010}.}

While discussions about AI welfare and moral patienthood understandably
focus on AI systems like robots and chatbots that appear conscious and
charismatic, many other AI systems --- like image generators or
algorithmic trading systems --- lack these features. Even if such
systems were in fact conscious and robustly agentic, we might not
recognize these capacities in them. And as these systems become
increasingly embedded in society, we might have increasingly strong
incentives to view them as mere objects. Companies, governments, and
other powerful actors who benefit from this technology might then
promote and reinforce our objectification of these systems, attempting
to frame moral consideration for these systems as fringe and unserious.

At present, it is an open question which kind of risk will be more
likely for particular kinds of AI systems, including seemingly conscious
and charismatic systems like robots and chatbots.\footnote{According to
  one survey, the majority of US residents sampled already endorse some
  chance that large language models might be conscious \citep{colombatto_folk_2024}.} The more advanced such systems become, the more
likely both risks might become in different respects: We might
over-attribute based on their behavioral similarities with humans, but
under-attribute based on their architectural differences from humans.
And the more economically dependent on chatbots we become, the more
likely over-attribution and under-attribution might become for them in
different respects as well: For example, we might over-attribute for
digital ``companions'' but under-attribute for other kinds of digital
minds.

While further research is required for a comprehensive assessment of
these risks, at least this much is plausible: Given our track record
with animals and the current pace of AI development, the risk of
under-attribution appears to be both reasonably likely and reasonably
harmful. To the extent that we also risk over-attribution, we cannot
simply avoid this risk by defaulting to treating AI systems as mere
objects. We should thus accept that AI welfare is difficult to get
right, and do the necessary work to improve our decisions --- by
assessing AI systems for evidence of consciousness, robust agency, and
other such capacities, and preparing policies and procedures for
treating AI systems with an appropriate level of moral concern.

\hypertarget{routes-to-near-term-ai-welfare}{%
\section{Routes to near-term AI
welfare}\label{routes-to-near-term-ai-welfare}}

\hypertarget{introduction-1}{%
\subsection{Introduction}\label{introduction-1}}

\textbf{We will argue that, according to the best information and
arguments currently available, there is a realistic possibility that
some AI systems will be moral patients in the near future.} We first
consider the possibility that some AI systems will be \emph{conscious}
in the near future, and we then consider the possibility that some AI
systems will be \emph{robustly agentic} in the near future.
Consciousness, robust agency, or both could suffice for moral
patienthood \emph{and} could exist in some near-future AI systems. In
our view, while these routes toward near-future AI moral patienthood are
far from certain, they are likely enough for AI companies to have a
responsibility to start implementing reasonable, proportionate
precautionary measures now.

We make structurally similar arguments for both routes. Each route
depends on a normative claim and a descriptive claim:

\textbf{Near-term consciousness: key claims}

There is a \textbf{realistic, non-negligible possibility} that:
\begin{enumerate}
    \tightlist 
    \item \textbf{Normative}: Consciousness suffices for moral patienthood,
    \emph{and}
    \item \textbf{Descriptive}: There are computational features --- like a
    global workspace, higher-order representations, or an attention schema
    --- that \emph{both}:
    \begin{itemize} 
        \tightlist 
        \item[a.] Suffice for consciousness, \emph{and}
        \item[b.] Will exist in some near-future AI systems.
    \end{itemize}
\end{enumerate}

\textbf{Near-term robust agency: key claims} 

There is a \textbf{realistic, non-negligible possibility} that:
\begin{enumerate}
    \tightlist 
    \item \textbf{Normative}: Robust agency suffices for moral patienthood,
    \emph{and}
    \item \textbf{Descriptive}: There are computational features --- like
    certain forms of planning, reasoning, or action-selection --- that
    \emph{both}:
    \begin{itemize}
        \tightlist 
        \item[a.] Suffice for robust agency, \emph{and}
        \item[b.] Will exist in some near-future AI systems.
    \end{itemize}
\end{enumerate}

In each case, the normative view is about the basis for moral
patienthood --- that is, about which capacities suffice for moral
patienthood. We still have substantial disagreement and uncertainty
about this issue. For example, are conscious experiences with a positive
or negative valence necessary for moral patienthood, or are conscious
experiences with a neutral valence sufficient? Similarly, is the ability
to set and pursue goals by rationally assessing your beliefs and desires
necessary for moral patienthood, or is the ability to set and pursue
goals by acting on your beliefs and desires sufficient? Experts continue
to debate such issues, but the kinds of consciousness and agency that we
consider here are among the leading views.\footnote{For some recent
  discussions of AI moral patienthood that review various criteria, see \cite{shevlin_how_2021,ladak_what_2024}. For a review of philosophical
  theories of moral patienthood, see \cite{jaworska_grounds_2021}.}

Meanwhile, in each case the descriptive view involves two parts. The
first part is about the basis for each of these capacities. For example,
are biological cells necessary for consciousness, or are digital chips
that play similar functional roles sufficient? Similarly, are beliefs
and desires with a propositional structure necessary for robust agency,
or are belief- and desire-like states that play similar functional roles
sufficient? Experts continue to debate these issues as well, and
especially in the case of consciousness, determining which features are
required for this capacity is widely regarded as one of the hardest
tasks in philosophy and science. We will thus draw from a range of
leading philosophical and scientific theories in our analysis.

The second part of each descriptive view is about the future of AI.
When, if ever, will these potentially morally significant features exist
in AI? Answering this question requires assessing both the pace and path
of AI development. Regarding the pace, will AI development slow down,
stay the same, or speed up? And regarding the path, it will be important
to consider two possibilities: (1) what we call \textbf{the direct
path}, which involves building conscious and/or agentic AI
\emph{intentionally}, because of their perceived intrinsic or
instrumental value; and (2) what we call \textbf{the indirect path},
which involves building conscious and/or agentic AI
\emph{unintentionally}, as a side effect of pursuing other, possibly
overlapping capabilities such as general intelligence.

Please note that in the following two sections we will mostly discuss
what we call the consciousness route and the robust agency route
separately, focusing on whether either one of these routes could lead to
AI welfare and moral patienthood on its own. However, we must also bear
in mind that these routes could also lead to this destination
\emph{together} --- that is, that there could be AI systems in the near
future with \emph{both} consciousness \emph{and} robust agency. And
however likely AI systems would be to morally matter for their own sakes
if they developed either of these capacities on its own, they would be
all the more likely to morally matter for their own sakes if they
developed both of these capacities together.

Finally, we emphasize that the aim of these arguments is not to
establish \emph{certainty} that these routes will lead to near-future AI
moral patients. The aim is instead to establish a \emph{realistic
possibility} that these routes will lead to this destination. The
details might differ in each case; for example, we might have more
confidence that consciousness suffices for moral patienthood but less
confidence that it will exist in near-future AI, whereas we might have
less confidence that robust agency suffices for moral patienthood but
more confidence that it will exist in near-future AI. But in each case,
as long as this route leads to a \emph{realistic possibility} of
near-future AI moral patients (in a sense of ``realistic possibility''
that we discuss below), the argument will be successful.

We focus on establishing a realistic possibility of near-future AI moral
patienthood for two reasons. First, the relevant normative and
descriptive issues are far too difficult and contested for anything
approaching certainty in either direction to be warranted at this stage.
Second, a realistic possibility of near-future AI moral patienthood is
all that we need for our purposes in this report, since that would
constitute a \textbf{morally significant risk} that merits consideration
now. Our conclusion in this report will thus simply be that AI companies
should start implementing low-cost, reasonable, proportionate steps to
consider and mitigate risks associated with AI welfare as we attempt to
improve our understanding of this topic over time.

Before we begin our discussion, a note about the scope of our discussion
in what follows. We will often discuss current large language models
(LLMs) due to their recent advances, current prominence, and salience in
AI welfare debates due to their conversational abilities. However, we
emphasize that one should not focus exclusively on current LLMs when
considering risks associated with near-future AI moral
patienthood.\footnote{For example, a recent TIME piece entitled ``No,
  Today's AI Isn't Sentient'' only discussed arguments against sentience
  in LLMs in particular, not AI systems more broadly \citep{li_no_2024}.} Many features that may be associated with
moral patienthood --- for example, embodiment, introspection, and
rationality --- are either already present in current non-LLM systems or
at least possible in near-future LLM or non-LLM systems, and we must
keep these possibilities in mind as well.

We now turn toward considering the two paths in more detail.

\hypertarget{consciousness-in-near-future-ai}{%
\subsection{Consciousness in near-future
AI}\label{consciousness-in-near-future-ai}}

The consciousness-based case for expecting moral patienthood in
near-term AI systems is that there is a \textbf{realistic,
non-negligible possibility} that:

\begin{enumerate}
    \tightlist 
    \item \textbf{Normative}: Consciousness suffices for moral patienthood,
    \emph{and}
    \item \textbf{Descriptive}: There are computational features --- like a
    global workspace, higher-order representations, or an attention schema
    --- that \emph{both}:
    \begin{itemize}
        \tightlist 
        \item[a.] Suffice for consciousness, \emph{and} 
        \item[b.] Will exist in some near-future AI systems.
    \end{itemize}
\end{enumerate}

We can now survey why each premise is plausible enough to support a
realistic possibility in near-future AI moral patienthood, given the
best information and arguments currently available.

\subsubsection{Does consciousness suffice for moral patienthood?}\label{subsubsec:does-consciousness-suffice-for-moral-patienthood}

The word ``consciousness'' is used in many different ways in ordinary
language and in various academic disciplines. In this report, we use
``consciousness'' to mean subjective experience --- what philosophers
call ``phenomenal consciousness.''\footnote{\cite{block_confusion_1995}} One famous way
of elucidating ``phenomenal consciousness'' is to say that an entity has
a conscious experience when there is ``something it is like'' for that
entity to be the subject of that experience.\footnote{\cite{nagel_what_1974}. For
  another way of elucidating the concept of consciousness via examples,
  in a way that seeks to be ``metaphysically and epistemically
  innocent'' with respect to philosophical assumptions, see \cite{schwitzgebel_phenomenal_2016}.} There is a subjective ``feel'' to your experiences as you
read this report: something that it is like to see the words on the
screen while, perhaps, listening to music playing through your speakers,
feeling the couch underneath you, feeling the laptop --- or a cat or a
dog --- on top of you.

The word ``sentience'' is likewise used in many different ways. Some
uses of ``sentience'' are synonymous with some uses of
``consciousness.'' But in this report, we use ``sentience'' to mean a
particular kind of consciousness, namely \emph{positively or negatively
valenced} conscious experiences. Anything that feels \emph{good} or
\emph{bad} in some way or another counts as positively or negatively
valenced in the relevant sense. This can include bodily states like
pleasures and pains and emotional states like hope and fear. If you find
the report engaging (or the opposite), if you find the music pleasant
(or the opposite), and if you find the couch comfortable (or the
opposite), then you are experiencing a range of positive or negative
states at the same time.

Why might sentience suffice for moral patienthood?\footnote{See, among
  others, \cite{singer_practical_2011,gruen2017conscious,nussbaum_frontiers_2007,dung_preserving_2024}.
  }
The idea that sentience is a sufficient condition for moral patienthood
is very plausible and widely accepted, because when you can consciously
experience positive and negative states like pleasure and pain, that
directly matters to you.\footnote{\cite{bentham_introduction_1789,rawls_theory_1971,degrazia_taking_1996,clarke_interest-based_2021}; \citet[p.~241]{parfit_what_2013}; \cite{korsgaard_fellow_2018}; \citet[p.~12]{kagan_how_2019}; \cite{roelofs_sentientism_2023,nussbaum_justice_2024}; \citet[chs.~2, 4]{birch_edge_2024}; \cite{smithies_hedonic_nodate}.
  } All else being equal, your life goes better for you when
you experience positive states like pleasure and your life goes worse
for you when you experience negative states like pain. So there is a
clear link between sentience and welfare. There is also a clear link
between sentience and moral patienthood, because we have a
responsibility not to harm welfare subjects unnecessarily, including and
especially by causing them to suffer unnecessarily.

To be clear, when we say that sentience suffices for moral patienthood,
we are not saying that sentience suffices for the specific kind of moral
status that typical adult humans possess.\footnote{ \cite{rawls_theory_1971}; \cite{nozick_anarchy_1974,korsgaard_fellow_2018,sebo_kantianism_2022}.} Typical adult humans are
\emph{both} sentient \emph{and} rational, which means that we have a
wide range of moral rights that these capacities jointly unlock (in
addition to having moral duties, though this is not our focus here). In
contrast, many other animals are plausibly sentient but non-rational,
which means that they plausibly lack certain moral rights, such as the
right to make their own medical decisions. But for present purposes,
what matters is that we at least have \emph{some} duties to animals,
including a duty to avoid causing them to suffer
unnecessarily.\footnote{\cite{korsgaard_fellow_2018}. See also \emph{A Theory of
  Justice} \citep{rawls_theory_1971}, which claims that, although animals are
  outside the scope of his theory, "the capacity for feelings of
  pleasure and pain and for the forms of life of which animals are
  capable clearly imposes duties of compassion and humanity in their
  case.''}

The idea that consciousness without valence suffices for moral
patienthood, while contested, is increasingly defended as
well.\footnote{See \cite{levy_moral_2009,chalmers_reality_2023,lee_consciousness_nodate,shepherd_consciousness_2018}.
  } Some philosophers argue for this
view by describing subjects who have consciousness without valence, and
by asserting that these subjects plausibly matter for their own
sakes.\footnote{Many of these arguments are discussed in \cite{ladak_what_2024}.}
However, some of these thought experiments describe subjects who have
consciousness \emph{and} agency, which makes it hard to tell whether
consciousness alone suffices.\footnote{For example, \cite{kagan_how_2019} argues
  that non-valenced consciousness \emph{with} preferences and desires is
  sufficient to warrant moral consideration. He imagines an entity with
  a preference for experiencing blue, though blue is not a valenced
  experience for the entity.} And while other thought experiments
describe subjects who have consciousness without valence \emph{or}
agency (say, a subject who passively experiences color with no pleasure
\emph{or} desire), the idea that this subject matters for their own sake
is more controversial.\footnote{For example, \cite{chalmers_reality_2023}
 argues that
  it seems wrong to kill a Vulcan, a hypothetical creature who is
  conscious but does not have valenced experience. While Vulcans can
  have desires, Chalmers argues that they would merit moral
  consideration even if they didn't have desires.}

In any case, even if consciousness is insufficient for moral patienthood
\emph{in theory}, it might still be sufficient --- or at least nearly
sufficient --- \emph{in practice}, since consciousness and valence might
be closely linked. For example, it might be that consciousness
necessarily involves valence,\footnote{\cite{cleeremans_consciousness_2022,lee_consciousness_nodate}}
in which case the emergence of conscious AI would suffice for the
emergence of sentient and morally significant AI. It might also be that
even if consciousness without valence is theoretically possible, the
step from conscious AI to sentient AI is much easier than the step from
non-conscious AI to conscious AI. In that case, the emergence of
conscious AI would be (at the very least) a significant step towards
morally significant AI, warranting careful scrutiny.\footnote{\cite{sebo_moral_2023}}

While this section is primarily about the consciousness route towards AI
welfare and moral patienthood, we note that consciousness could be
relevant for other reasons too; for instance, as noted above, whether or
not consciousness suffices for moral patienthood on its own, it might
suffice in combination with other capacities like robust agency --- that
is, it might be that when entities can consciously set and pursue their
own goals based on their own beliefs and desires, they matter for their
own sakes (whether or not they can experience pleasure or pain). If so,
then the emergence of conscious AI would increase the probability of
morally significant AI for this reason (that is, as part of a
consciousness \emph{and} robust agency route) as well.

In the remainder of this section we follow our strategy in \cite{sebo_moral_2023} by examining the route towards conscious AI in general rather
than examining the routes towards specific kinds of consciousness, such
as valenced consciousness. We make this choice for the sake of
simplicity and specificity, given that the science of consciousness is
more developed than the science of valence and given that consciousness
is plausibly either sufficient for or otherwise closely linked to
valence and/or moral patienthood anyway. We will then, in upcoming work,
discuss these capacities in more detail and present a research agenda
that examines how to identify indicators for both consciousness and
sentience in LLMs and other AI systems.

\subsubsection{Will some AI systems be conscious in the near future?}\label{subsubsec:will-some-ai-systems-be-conscious-in-near-future}

How can we tell whether AI systems are conscious? We can be confident
that other (awake, adult) humans are conscious, since each of us knows
that \emph{we} are conscious and that other humans are behaviorally and
anatomically similar to us. But we have uncertainty about \emph{why} we
are conscious, that is, about which features of our brains or bodies are
responsible for, or associated with, consciousness. We also have
uncertainty about which \emph{other animals} are conscious,\footnote{This
  is sometimes called the ``distribution question'' (\cite{allen_animal_2000}.}
because of uncertainty not only about which features are associated with
consciousness in humans, but also about how to extrapolate what we know
about human consciousness to the nonhuman animal case. We also have
significant uncertainty about how many animals' brains work.

Recently, scientists have attempted to improve our understanding of
nonhuman consciousness by searching for what ``markers'' of
consciousness in other animals.\footnote{See \cite{allen_animal_2007}.
  For philosophical defenses see \cite{tye_tense_2016,birch_review_2021,bayne_bayne_2021}
  For scientific use of markers see \cite{braithwaite_fish_2010,sneddon_defining_2014,birch_review_2021} 
  .} At a high level this
method proceeds as follows: We start by distinguishing between certain
kinds of conscious and unconscious processing in humans --- say,
distinguishing \emph{pain} from \emph{nociception}\footnote{Nociception
  is the physiological processing of noxious, or harmful, stimuli. It may or may not
  be accompanied by the qualitative experience of pain.} by seeing when
patients do or do not report consciously feeling pain. We then identify
features that correlate with conscious processing --- say, certain
behaviors, brain regions, or patterns of neural processing. We then
search for relevantly similar features in nonhumans, and we treat the
presence of these features as \emph{evidence} of conscious processing.

This method does not tell us which animals definitely are or are not
conscious, and there are a variety of methodological difficulties
related to identifying and extrapolating the relevant
features.\footnote{Most saliently, what we learn from the human case is,
  most directly, what some of the \emph{sufficient} conditions are for
  consciousness. It is difficult to know which elements we should take
  to be strictly necessary.} But this method has still allowed
researchers and policymakers to make more informed estimates and
decisions about animal welfare despite ongoing disagreement and
uncertainty about animal consciousness. Moving forward, the same can be
true for AI. Later on, we discuss how to tailor this method for AI, but
for now, we can emphasize that it will be important to focus less on
\emph{behavioral} evidence (with a limited class of exceptions, which we
discuss below) and more on \emph{internal} evidence, like architectural
and computational features.

Which kinds of architectural and computational features might indicate
consciousness in AI? We can look to \textbf{neuroscientific theories of
consciousness} for guidance. These theories use a variety of empirical
methods to uncover which states and processes are associated with
consciousness.\footnote{\cite{seth2022theories}; \cite{chalmers_how_2004}}
While
these theories tend to be framed around \emph{brain} or \emph{neural}
states and processes (given our typical focus on humans and other
animals), the ones we focus on tend to specify these states and
processes in terms of the \emph{computations} that they perform or the
\emph{functional roles} that they play. While these theories remain
contested and incomplete, they can still shed light on what kinds of
states and processes might be associated with consciousness in AI.

Of course, even if we identify a variety of architectural and
computational markers of consciousness in AI systems, we must still ask
whether these markers \emph{suffice} for consciousness.
\textbf{Computational functionalism} is the hypothesis that some class
of computations suffices for consciousness.\footnote{Computational
  functionalism is one kind of functionalism \citep{piccinini_non-computational_2018}. For a
  survey of varieties of functionalism, see \cite{block_comparing_2009,maley_get_2013}. See \cite{putnam_psychological_1967} for a classic statement of computational functionalism. See \cite{colombo_computational_2023} and \cite{rescorla_computational_2020} for computational theories of mental
  phenomena more generally.} If this hypothesis is correct, then the
question is which computations suffice for consciousness and when, if
ever, these computations will exist in AI.\footnote{Note that the
  theories of consciousness we consider, and the investigations of AI
  systems that we propose, do not purport to solve the ``hard problem''
  of consciousness, which concerns how physical processes relate to
  conscious experiences. Whatever mysteries there may be about this
  fundamental issue, virtually everyone agrees that physical processes
  (such as certain patterns of neural firing or certain computations)
  and conscious experiences are closely related in systematic ways. The
  theories and investigations at hand seek to find the neural processes
  and/or computations that are associated with consciousness.} If this
hypothesis is incorrect\footnote{\cite{godfrey-smith_other_2016,godfrey-smith_metazoa_2020,cao_multiple_2022,seth_being_2021,seth_conscious_2024}}, then AI consciousness will remain at best
a theoretical possibility until we move beyond current architectures. 
For example, if more biology-like functions are required for consciousness, then AI consciousness may require novel hardware that can perform those functions.\footnote{\cite{brunet_minds_2020}}

As we discuss below, our view is that \textbf{computational
functionalism is neither clearly correct nor clearly incorrect} at this
stage. We might lean one way or the other, but given the importance and
difficulty of consciousness as a research topic, we should leave room
for doubt. That means that AI consciousness assessments will need to be
probabilistic rather than all or nothing at present and, plausibly, for
the foreseeable future. To use a simple example, if we estimate that
there is a 30--50\% chance that computational functionalism is correct
and a 30--50\% chance that an AI system is conscious if so, then it
follows that there is a 9--25\% chance that this AI system is conscious.
That would be good to know when interacting with this AI system!

In this precautionary spirit, some of the authors of this report
(Patrick Butlin, Robert Long, and Jonathan Birch) released a paper in
2023 exploring the implications of several prominent scientific theories
of consciousness\footnote{\cite{butlin_consciousness_2023}} --- viewed through the
lens of computational functionalism --- for AI consciousness. The table
below lists the theories and conditions we surveyed:

\begin{longtable}{|>{\arraybackslash}p{\textwidth}|}
\hline 
\textbf{Recurrent processing theory} \\
\hline 
1.1 Input modules using algorithmic recurrence \\
\hline 
1.2 Input modules generating organised, integrated perceptual
representations \\
\hline 
\textbf{Global workspace theory} \\
\hline 
2.1 Multiple specialised systems capable of operating in parallel
(modules) \\
\hline 
2.2 Limited capacity workspace, entailing a bottleneck in information
flow and a selective attention mechanism \\
\hline 
2.3 Global broadcast of information in the workspace to all modules \\
\hline 
2.4 State-dependent attention, giving rise to the capacity to use the
workspace to query modules in succession to perform complex tasks \\
\hline 
\textbf{Computational higher-order theories} \\
\hline 
3.1 Generative, top-down or noisy perception modules \\
\hline 
3.2 Metacognitive monitoring distinguishing reliable perceptual
representations from noise \\
\hline 
3.3 Agency guided by a general belief-formation and action selection
system, and a strong disposition to update beliefs in accordance with
the outputs of metacognitive monitoring \\
\hline 
3.4 Sparse and smooth coding generating a `quality space' \\
\hline 
\textbf{Attention schema theory} \\
\hline 
4.1 A predictive model representing and enabling control over the
current state of attention \\
\hline 
\textbf{Predictive processing} \\
\hline 
5.1 Input modules using predictive coding \\
\hline 
\textbf{Agency and embodiment} \\
\hline 
6.1 Minimal agency, that is, the capacity to learn from feedback and
select outputs in such a way as to pursue goals, especially involving
flexible responsiveness to competing goals \\
\hline 
6.2 Embodiment, that is, the capacity to model output-input
contingencies, including some systematic effects, and to use this model
in perception or control \\
\hline 
\end{longtable}

These theories come in numerous versions. Weaker versions are directed only at distinguishing conscious from unconscious states in humans. For our purposes, what matters are stronger versions that aim to give sufficient conditions for consciousness across human and nonhuman systems. The crucial claim is that there is a realistic possibility that such a theory is correct, at least to the extent that the indicators associated with one or more theories jointly provide sufficient conditions for consciousness. Given this, and given that the indicators can be implemented in near-term AI systems, then there is a realistic possibility that near-term AI systems will be conscious.

It is by no means obvious that any of these theories are correct, especially in their stronger versions. Our claim is not that they are correct, but only that there is a realistic possibility that one of them is correct. In fact, for our purposes, it would suffice if an extension of one of these theories is correct, or if some other computational theory is correct, as long as the relevant theory provides sufficient conditions that can be implemented in near-future AI.

After surveying the indicators and a variety of AI systems and methods,
we found no clear barriers to satisfying these indicators using current
AI architectures and methods. And while no current systems seemed very
likely to be conscious at the time, there do seem to be plausible routes
towards conscious AI, according to these theories and assumptions.

Consider global workspace theory, which associates consciousness with a
global workspace --- roughly, a system that integrates information from
mostly-independent, task-specific information-processing modules, then
broadcasts it back to them in a way that enables complex tasks like
planning. We found that, as argued by \cite{juliani_link_2022}, some AI
architectures already embody some, though not all, aspects of a global
workspace.\footnote{Similarly, \cite{goldstein_ai_nodate} have
  argued that language agents instantiate a global workspace in the
  relevant sense.} We also found that several technical research
programs seem poised to implement further aspects of a global workspace
in the near future.\footnote{\cite{vanrullen_deep_2021,goyal_retrieval-augmented_2022,goyal_inductive_2022,bao_multimodal_2020}}
   And indeed, researchers
investigating consciousness in AI recently have subsequently built a
system which aims to implement all of the global workspace indicators
from \cite{butlin_consciousness_2023}.\footnote{\cite{dossa_design_2024}} We found similar
trends for other theories as well.\footnote{See \cite{butlin_consciousness_2023}, \cref{recommendations-for-ai-companies}.}

For the global workspace and other potential indicators of
consciousness, progress might continue via direct efforts, in which
researchers try to build conscious AI \emph{intentionally}. Some
researchers are motivated to build conscious AI because they view
conscious AI as an end in itself. ``It would be monumentally
cool,'' as \citet[p.~7]{graziano_attention_2017} puts it. Others
are motivated to build conscious AI because they believe that
consciousness or related features could make AI systems safer or more
capable.\footnote{\cite{bengio_system_2019,goyal_inductive_2022,graziano_attention_2017,verschure_synthetic_2016}}

AI companies might also build conscious AI \emph{unintentionally}. You
might wonder why an AI system designed to, say, navigate a warehouse or
manage a factory would be conscious. The answer is that on some views of
consciousness, the building blocks of consciousness can emerge as a side
effect of other cognitive capacities, such as perception, cognition, and
robust agency. For example, many theories of consciousness take
\emph{metacognition} --- roughly, the ability to model your own
cognition --- as important for both consciousness and
decision-making.\footnote{\cite{shea_supra-personal_2014}. For an overview of
  higher-order theories of consciousness, see \cite{carruthers_higher-order_2020}. As David Chalmers has noted (\cite{chalmers2018meta}), ``In general, we should expect
any intelligent system to have an internal model of its own cognitive
states.'' } 
This kind of connection is
also an element of, among others, global workspace theory,\footnote{\cite{shea_global_2019}} higher-order theories, and the attention schema
theory.

The link between consciousness and other cognitive capacities is
unclear, and we certainly do not want readers to naively equate them.
But it may be that increased cognitive capacity tends to bring about
consciousness, in both biological and digital systems. Thus, it may even
be that general intelligence will itself be a plausible indicator for AI
consciousness in the future.\footnote{\cite{shevlin_general_2020}. A related view is \citeauthor{danaher_welcoming_2020}'s (\citeyear{danaher_welcoming_2020}) ``ethical behaviorism,'' which holds that we should
  treat an entity as a moral patient if it is ``performatively
  equivalent to other entities that have significant moral status.''
  Danaher stresses equivalence of behavior rather than equivalence of
  cognitive capacities.}

For these and other reasons, we agree with a recent open letter by
consciousness scientists and AI researchers that ``it is no longer in
the realm of science fiction to imagine AI systems having feelings and
even human-level consciousness.''\footnote{\cite{amcs_responsible_2023}} As one of the
authors of this report, David Chalmers, wrote in 2023:

\begin{quote}
I think it wouldn't be unreasonable to have a credence over 50 percent that we'll
have sophisticated LLM+ systems (that is, LLM+ systems with behavior that 
seems comparable to that of animals that we take to be conscious) 
with all of these properties\footnote{With ``all
  of these properties,'' Chalmers is referring to a number of proposed
  necessary conditions for consciousness, quite similar to those
  discussed here, that he argues will plausibly be satisfied by
  near-term AI systems: senses, embodiment, world-models, self-models,
  recurrent processing, global workspace, and unified agency.}
\textbf{within a decade}. It also wouldn't be unreasonable to have at least a
50 percent credence that if we develop sophisticated systems with all of these
properties, they will be conscious. \textbf{Those figures would leave us
with a credence of 25 percent } \textbf{or more} [emphasis
ours].\footnote{\cite{chalmers_could_2023}. Chalmers also notes that these credences follow from
  mainstream views about consciousness; he adds that his own credences
  are higher, given that his own views about consciousness are more
  expansive.}
\end{quote}

\hypertarget{robust-agency-in-near-future-ai}{%
\subsection{Robust agency in near-future
AI}\label{robust-agency-in-near-future-ai}}

The robust agency-based case for expecting moral patienthood in
near-term AI systems is that there is a \textbf{realistic,
non-negligible possibility} that:

\begin{enumerate}
    \tightlist 
    \item \textbf{Normative}: Robust agency suffices for moral patienthood,
    \emph{and}
    \item \textbf{Descriptive}: There are computational features --- like
    certain forms of planning, reasoning, or self-awareness --- that
    \emph{both}:
    \begin{itemize}
        \tightlist 
        \item Suffice for robust agency, \emph{and} 
        \item Will exist in some near-future AI systems.
    \end{itemize}
\end{enumerate}

We will present arguments for each of these premises in turn.

\subsubsection{Does robust agency suffice for moral patienthood?}\label{subsubsec:does-robust-agency-suffice-for-moral-patienthood}

The word ``agency'' is used in many different ways as well. In a broad
sense, one might use ``agent'' to mean any entity that senses the
environment and responds,\footnote{\citet[p.~34]{russell_artificial_2010}:
  ``an agent is anything that can be viewed as perceiving its
  environment through sensors and acting upon that environment through
  activators.''} which would include thermostats, or any entity that
learns and pursues goals, which would include very simple RL agents that
play Tic-Tac-Toe. Some might take even these basic ways of being an
agent to be sufficient for moral patienthood and, to the extent that you
accept this view, you likely already endorse moral consideration for AI
systems without needing to read further. While this view merits
consideration (and we consider it further in upcoming work), we set it
aside for now. We will be arguing that AI systems could be agents in a
more demanding sense.\footnote{\cite{wooldridge_intelligent_1995,schlosser_agency_2019,kenton_discovering_2023}}

Specifically, ``robust agency'' is the ability to pursue goals via some
particular set of cognitive states and processes. Which ones? There are
several ``levels'' of agency that extend beyond the mere ability to
learn and pursue goals, and that could plausibly suffice for moral
patienthood even when consciousness is absent.\footnote{For another
  taxonomy of agency for the AI context, see \cite{dung_preserving_2024}, who also notes
  the importance of agency for AI moral patienthood.} For present
purposes, we highlight three such levels:

\begin{enumerate}
\item
  \textbf{Intentional agency}: This is the capacity to set and pursue
  goals via beliefs, desires, and intentions. Roughly, if you have
  mental states that represent \emph{what is}, \emph{ought to be}, and
  \emph{what to do}, and if these mental states work together in the
  right kind of way to convert perceptual inputs to behavioral outputs,
  then you count as an intentional agent.
\item
  \textbf{Reflective agency}: This is intentional agency plus the
  ability to \emph{reflectively endorse} your own beliefs, desires, and
  intentions. Roughly, if you can form beliefs, desires, and intentions
  about your own beliefs, desires, and intentions, accepting or
  rejecting your own attitudes and behaviors at a higher level, then you
  count as a reflective agent.
\item
  \textbf{Rational agency}: This is reflective agency plus the ability
  to \emph{rationally assess} your own beliefs, desires, and intentions.
  Roughly, if you can consider whether particular beliefs, desires,
  intentions, actions are justified and adopt principles that you can
  treat as rules of conduct, then you count as a rational
  agent.\footnote{For any kind of agency, one key question will be
    whether that form of agency requires consciousness. For example,
    when assessing whether an AI system has beliefs and desires, one
    view would be that it is enough to have states that play the
    functional role of beliefs and desires (roughly, representing what
    is and what ought to be). On this view, beliefs and desires would
    not require consciousness. But another view would be that beliefs
    and desires might require that it feels like something for the AI
    system to perform those functions --- roughly, it needs to have
    conscious feelings about how the world is and ought to be. On this
    view, beliefs and desires would require consciousness. For an
    overview of phenomenal intentionality, see \cite{bourget_phenomenal_2019}.}
\end{enumerate}

There might also be different ways of realizing each capacity, with
different kinds and degrees of cognitive sophistication. For example,
typical adult humans have the capacity for propositional thought, which
means that our thoughts can have a structure that allows for truth
values and logical relations. This capacity unlocks powerful forms of
intentional agency, reflective agency, \emph{and} rational agency, since
it allows us to develop a wide range of novel beliefs, desires, and
intentions and then use evidence and reason to assess their accuracy and
coherence. In contrast, while nonhuman animals appear to lack this
capacity, many animals at least have a limited capacity for symbolic
thought, metacognition, and planning and problem solving.\footnote{For
  example, many animals appear to have thoughts that take the form of
  maps and charts, with abstract information about social and
  environmental structures. Many animals also have perceptual
  affordances that represent what is, what ought to be, and what to do;
  a limited ability to represent their own mental states and the mental
  states of others; and a limited ability to make basic inferences based
  on disjunctions and negations. These capacities allow for a limited
  form of all three levels of robust agency, with important similarities
  with and differences from the typical human form. For discussion, see \cite{sebo_agency_2017,camp_language_2009,bermudez_mindreading_2009,gennaro_animals_2009,lurz_self-awareness_2009}.}

Why might these kinds of agency suffice for moral patienthood? First,
intentional agents are potentially \emph{welfare subjects}.\footnote{This
  kind of view reflects one of the main philosophical theories of
  welfare, known as the desire-satisfaction view. This view holds that
  your life goes better or worse for you to the extent that your desires
  are satisfied or frustrated, independently of whether you consciously
  experience the satisfaction or frustration of your desires. See, for
  instance, \cite{heathwood_desire-fulfillment_2015}.} It seems plausible that when you have
desires, your life goes better for you when your desires are satisfied
and worse for you when your desires are frustrated. Moreover, the
satisfaction or frustration of desires can benefit or harm you
\emph{whether or not} you consciously experience them.\footnote{\cite{dorsey_desire-satisfaction_2013}} On some views, this is why we can be posthumously harmed, for
example.\footnote{See, for example, \cite{nozick_anarchy_1974,rachels_end_1986,kagan_me_1994,ruddick_biographical_2005}.} And while some people think that
desire-satisfaction and desire-frustration matter only for conscious
beings, others think that they matter for non-conscious beings as well,
and so animals with desires deserve moral consideration whether or not
they can consciously experience pain, for instance.\footnote{See \cite{neely_machines_2014,kagan_how_2019,kammerer_ethics_2022,delon_against_2024,goldstein_ai_nodate}, who consider and/or defend versions of
  this view.}

Reflective agency, particularly in its propositional form, then adds the
ability to have desires about our own desires, which is at the root of
some conceptions of free will, the self, and personal
identity.\footnote{\cite{frankfurt_freedom_1971}} Specifically, our desires become
``ours'' in a new sense when we endorse them through reflection.
Reflective agency also allows for new kinds of morally significant
interests and relationships. When you can have mental states about other
mental states, you can have at least a limited conception of how you and
others think or feel. For example, reflective agents can have
preferences about how they relate to each other, and all else being
equal, their lives are better for them when those preferences are
satisfied and worse for them otherwise.

Rational agency, particularly in its propositional form, then adds the
ability to create \emph{social contracts} with other rational agents
(assuming that they can communicate as well), which are at the root of
some conceptions of moral, legal, and political rights and
responsibilities.\footnote{Traditionally, philosophers have focused more
  on the significance that these capacities might have for \emph{moral
  agency}. This is a different concept from moral patienthood: In
  general, you are a moral agent if you can have duties to others, and
  you are a moral patient if others can have duties to you. And many
  philosophers believe that a particular kind of agency --- namely,
  rationality --- is important for moral agency because you need to be
  able to reason about your actions in order to be morally accountable
  or responsible for them. When a non-rational entity (say, a hurricane)
  kills someone, we can call it bad. But when a rational agent knowingly
  and willingly kills someone, we also call it wrong.} Rational agency
also allows for decisions based on judgments about reasons and
principles, and this ability not only allows for new kinds of interests,
but also --- on some views --- commands a kind of respect that extends
beyond compassion. Indeed, this idea of respect for rational agents is
at the root of the Kantian ethical theory, which rests alongside the
utilitarian idea of beneficence for sentient beings as one of the two
most influential ethical theories in the modern era.\footnote{However,
  please note that while Immanuel Kant (1785) accepted rationality as
  the basis for moral patienthood, some contemporary Kantians, such as \cite{korsgaard_fellow_2018},
  now accept sentience as the basis for
  moral patienthood instead.}

In what follows, we discuss the route towards robust agency in general
rather than the routes towards intentional, reflective, and rational
agency in particular. We distinguish these levels of robust agency here
to emphasize that there can be different kinds of robust agency with
different kinds of moral significance, both within and across these
levels. That means that when we search for robust agency in nonhumans,
including animals and AI systems, it would be a mistake to anchor too
much on human agency. But having now made this point, we focus on
showing that AI development is currently on a path to create many
computations that we associate with all of these levels of robust
agency.

\subsubsection{Will some AI systems be robustly agentic in the near
future?}\label{subsubsec:will-some-ai-systems-be-robustly-agentic}

There are plausible routes by which we might soon have AI systems with
robust agency. What it takes to be an intentional, reflective, and/or
rational agent is not clear. Many cognitive capacities are plausibly
associated with each level --- capacities that are the aims of
well-resourced research programs that are making significant progress.
For example, intentional agency is related to capacities for planning,
memory, and learning. Reflective agency is related to introspection and
situational awareness. And rational agency is related to all of these
capacities, along with abstract reasoning about principles and
strategies.

The capacities described in this section might or might not be necessary
or sufficient for these levels of robust agency. But to the extent that
these capacities are present, the probability of robust agency will
increase. And the development of AI systems with such capacities aligns
with the aims of significant research and development efforts: Both
major tech companies and startups are investing heavily in creating more
agentic AI systems, with significant progress already
evident.\footnote{\cite{toner_through_2024}} In what follows, we discuss recent
work in reinforcement learning, language agents, and other research
programs as case studies.\footnote{We focus on the research programs
  that seem in recent years to be making the most strides, though
  doubtless there have been and likely will be other paradigms that make
  progress.}

First, \textbf{reinforcement learning (RL)} is the subfield of AI most
concerned with building agents as a fundamental goal, as Sutton and
Barto put it in their canonical RL textbook:

\begin{quote}
[RL] explicitly considers the whole problem of a goal-directed agent
interacting with an uncertain environment.\footnote{\citet[p.~3]{sutton_reinforcement_2015}}
\end{quote}

In RL, the aim is to write algorithms that allow agents to learn,
reason, and act in pursuit of specified goals in complex environments.
RL construes goal-pursuit as maximizing reward through interaction with
the environment, and some RL researchers argue that this process allows
agents to acquire the whole suite of capacities observed in intelligent
systems.\footnote{\cite{silver_reward_2021}}

RL is a useful starting point for examining robust agency in AI, not
only because of its prominence and recent successes, but also because of
its centrality of reinforcement learning for human and nonhuman
agency.\footnote{\cite{dolan_goals_2013}} Of course, in humans and many
other animals, a rich understanding of social and environmental context
and the expressive power of language, among other capabilities, make
substantial contributions to our capacities for robust agency as well.
But deep RL has made it possible for AI agents to be virtually embodied
and situated in environments comparable to those inhabited by
animals,\footnote{\cite{shanahan_artificial_2020,abramson_imitating_2021}} and so
it may be a compelling foundation for projects to emulate natural
agency.

Deep RL has achieved significant successes in game-playing in the last
decade. These successes include superhuman performance in Go, chess,
shogi, and a variety of Atari games (i.e. MuZero),\footnote{\cite{schrittwieser_mastering_2020}} as well as in games that involve controlling an avatar
in a relatively rich and dynamic environment, such as Gran Turismo (GT
Sophy)\footnote{\cite{wurman_outracing_2022}} and Starcraft II
(AlphaStar).\footnote{\cite{vinyals_grandmaster_2019}} Many of these agents, such as
MuZero, can learn models of the game environments (including opponent
behavior) and use them to make the predictions in order to plan. Some
game-playing agents can also exploit a form of episodic memory, called
experience replay, in order to increase sample efficiency.\footnote{\cite{mnih_human-level_2015}}

While many agents have relatively narrow capabilities, DeepMind's
Adaptive Agent demonstrates the ability to rapidly adapt to
new tasks in a 3D virtual environment.\footnote{\cite{adaptiveagentteam2023humantimescaleadaptationopenendedtask}} AdA is trained on a varied
curriculum of tasks, inducing \emph{meta-learning} of an algorithm for
few-shot learning of new tasks --- that is, for learning how to make
reliable predictions and decisions based on a small number of examples.
The architecture includes a Transformer-based memory module encoding
recent observations, allowing the system to identify dependencies
between actions and subsequent events. As a result, AdA has a notably
effective, flexible way of acquiring and using a grasp of the
environment's dynamics.

RL research targets robust agency partly through research on abstraction
and hierarchical planning. In complex environments that require
intricate sequences of movements, it can be vital to use representations
that abstract away from low-level details. For example, Director can
learn to break down tasks with sparse rewards into subgoals.\footnote{\cite{hafner_deep_2022}} RL research also targets robust agency through research
on multiplayer strategy games such as Diplomacy, which involves forming
alliances. For example, Meta's Cicero\footnote{\cite{meta2022human}}
uses an adapted language
model to achieve comparable performance to high-level human players,
planning in ways that predict human behavior and changing plans through
communication with these other agents.

Researchers are now pursuing several promising strategies involving
\textbf{language agents} as well.\footnote{\cite{mialon_augmented_2023,wang_survey_2024,sumers_cognitive_2024,guo_large_2024}} Language agents
leverage the powerful natural language processing and generation
abilities of LLMs for greater capability and flexibility, by embedding
LLMs within larger architectures that support functions like memory,
planning, reasoning, and action selection.\footnote{Depending on one's
  definition, even LLMs that are used as (or in) chatbots, like ChatGPT,
  may also qualify as ``language agents'' (see \cite{butlin_agency_nodate} for
  discussion). \cite{goldstein_does_2024} argue that such LLMs
  also plausibly have beliefs and desires, noting the importance of this
  claim for issues of moral patienthood.} While existing systems
struggle with reliability, the properties of language agents and their
initial success suggest that they have the potential to overcome
traditional barriers towards more agentic systems\footnote{LLMs in their own right are also making progress towards agency, and in particular when combined with RL methods. Recent work has explored using language models as the starting point for RL training, as in reinforcement learning from human feedback (RLHF), LM-based proof-writing agents, and LM-based coding agents (\cite{ruan2024observationalscalinglawspredictability}). Indeed recent benchmarking work suggests progress on language modeling capabilities naturally results in performance gains on agentic tasks. See \cite{ziegler2020finetuninglanguagemodelshuman}; \cite{DeepMindIMO}; \cite{gehring2024rlefgroundingcodellms}}. Indeed, several
notable examples of language agents have emerged in recent years,
demonstrating that this strategy can lead to more robust and generalized
agency across diverse domains:\footnote{See \cite{butlin_agency_nodate}, who discusses
  several of these systems in light of various conceptions of agency.}
  
\begin{itemize}
\item
  ReAct \citep{yao_react_2023} alternates between generating thoughts/plans
  and taking actions in interactive environments. It can break down
  complex tasks, gather information dynamically, and adjust its approach
  based on intermediate results. ReAct has shown strong performance on
  language-based tasks like question answering and web navigation.
\item
  Generative Agents \citep{park_generative_2023} simulates interactive AI
  characters. The agents have persistent identities, relationships, and
  goals, with an LLM generating plans and actions based on their
  memories, observations, and reflections. As a result, they exhibit
  long-term coherence with evolving goals and emergent social behaviors.
\item
  Voyager \citep{wang_voyager_2023} uses an LLM to control an embodied agent
  in Minecraft, iteratively setting its own goals, devising plans, and
  writing code to accomplish increasingly complex tasks. By maintaining
  a skill library and reflecting on past experiences, Voyager can
  bootstrap its way to mastering the game\textquotesingle s tech tree
  and creatively solving novel challenges.
\item
  SayCan \citep{ahn_as_2022} grounds language in robotic control, using
  an LLM to generate high-level plans that are then mapped to concrete
  robot skills. This allows the system to flexibly respond to natural
  language commands by reasoning about affordances and breaking tasks
  into actionable steps.
\end{itemize}

These language agents, and others, exhibit several key properties that
make them more robustly agentic compared to many traditional AI systems
(though progress has certainly been made outside the language agent
paradigm as well):

\begin{itemize}
\item
  Flexible goal-setting and planning: Rather than being constrained to
  predefined reward functions, language agents can understand open-ended
  objectives, generate their own subgoals, and devise multi-step plans
  to achieve them.
\item
  Adaptive reasoning: By leveraging LLMs\textquotesingle{} broad
  knowledge and reasoning capabilities, language agents can navigate
  novel contexts, drawing from relevant insights in other contexts to
  inform their decisions.
\item
  Memory integration: Many language agents incorporate episodic and
  semantic memory systems, allowing them to learn from experience,
  maintain consistent behaviors, and apply past knowledge to new
  contexts.\footnote{cf. also, among others, \cite{rubin_learning_2022,shinn_reflexion_2023}.}
\item
  Metacognition: Agents like Voyager and Generative Agents can reflect
  on their own thoughts and experiences, enabling higher-order reasoning
  and self-improvement.
\item
  Open-ended interaction: These systems can often engage in natural
  language dialogue, explain their reasoning, and incorporate new
  information or instructions on the fly.
\end{itemize}

For these and other, similar reasons, \cite{goldstein_ai_nodate} argue that philosophical consideration of language agents
suggests that ``the technology already exists to create AI systems with
wellbeing.'' Our conclusion at this stage is somewhat more tentative:
Rather than assert or deny that the technology already exists to create
such systems, we merely assert that the technology already exists to
create \emph{key properties} of such systems.\footnote{Language agents
  and other generative agents are clearly quite different from
  biological agents. For example, their `observations' are all in
  language, their `intelligence' is mostly embedded within language
  models, and many of them do not learn via reward and punishment. It is
  hard to know exactly what to say about them, and we are not claiming
  that they have robust agency. Instead, we are claiming that they
  constitute a significant step towards robust agency in AI. While
  language agents and other generative agents are still limited in many
  ways, their ability to flexibly pursue goals, reason about abstract
  concepts, and adapt to novel situations is a striking indicator of
  what else may soon be possible.} We also note that, while many
``language agents'' may turn out to be impressive demos that do not
scale, frontier language models are also being made more agentic by the
day\footnote{\cite{anthropic_introducing_2024}}. The question is whether and to what extent progress will continue
from here. We believe that the current state of the field, combined with
the clear incentives that developers have for continuing to build
towards robust agency, strongly suggest that progress will continue.

Specifically, we expect that \textbf{future routes} have the potential
to produce further properties associated with robust agency. Consider an
AI that combines the learning capabilities of RL agents with the
reflection and world modeling capabilities of LLMs. In a recent
interview, Demis Hassabis discusses one possible example of such a
system: combining LLMs with the Monte Carlo Tree Search (MCTS) used by,
say, AlphaGo.\footnote{\cite{fridman_lex_2022}} In such a system, the LLM
provides a rich, flexible "belief" system about the world. The LLM could
be used to analyze the system\textquotesingle s decision-making process,
approaching a form of meta-cognition. This system might also be able to
provide explanations or justifications for its decisions, approaching a
form of rational deliberation.\footnote{This is just one kind of
  potential capability enhancement from synergies between LLMs and RL;
  \cite{pternea_rlllm_2024} surveys a variety.}

Applying RL to augment run-time performance of LLMs is a promising route
as well. Consider OpenAI's o1 system. Described as a "Large Reasoning
Model" (LRM), o1 incorporates extensive chain-of-thought reasoning,
which is speculated to be trained using RL. This approach, combined with
increased compute usage at inference time, has led to markedly improved
performance on planning and reasoning tasks, nearly saturating some
benchmarks that previously challenged LLMs. o1 also exhibits concerning
behaviors related to instrumental convergence, deceptive alignment, and
reward hacking. These traits, while currently limited in scope and
impact, highlight the potential for LLMs to be used to make systems that
are more agentic.

The long-term ambitions of AI research have consistently aimed towards
creating systems that exhibit key characteristics of robust agency.
These goals include developing AI systems that can operate on extended
time horizons, maintain coherent objectives, engage in self-reflection,
and revise their own goals and methods. For large swaths of the field,
the ultimate aim has been and remains to create "human-level" AI capable
of general problem-solving across diverse domains.\footnote{See, among
  many others, \cite{mccarthy_proposal_1955,moravec_mind_1995}, \citet[ch.~27]{russell_artificial_2010}, \cite{morris_levels_2024}.} This vision implicitly
requires many of the capacities we associate with robust agency:
intentional goal-setting, reflective self-assessment, and rational
decision-making in complex, dynamic environments.

These aims are also central to current investment and effort. For
example, major tech companies like Microsoft and Google have announced
plans for AI tools with ``more autonomy and less human intervention''
and agents that can autonomously carry out complex multi-step
tasks.\footnote{Microsoft touts systems ``that can now act as
  independent agents --- ones that can be triggered by events --- not
  just conversation --- and can automate and orchestrate complex,
  long-running business processes with more autonomy and less human
  intervention.'' See \cite{aftab_microsoft_2024}.} Simultaneously, startups such as
Adept, MultiOn, and Lindy have raised hundreds of millions of dollars to
develop flexible AI agents. This excitement could be overstated and/or
misplaced, of course --- and recent work has already seen periods of
hype and disillusionment, as with systems like AutoGPT --- but it is
evidence that a lot of effort will be put into building more agentic AI
systems in the coming years.

While today\textquotesingle s dominant AI paradigm, centered around
LLMs, often appears less explicitly agentic, there are clear signs of a
shift towards more agent-like systems like the kinds discussed here.
Consider the goal of building advanced AI assistants that can undertake
a variety of complex tasks.\footnote{See \cite{gabriel_ethics_2024,dong_towards_2023}} Such a system would need to be able to set and pursue
long-term goals based on high-level reflections about facts and
values.\footnote{\cite{lecun_objective-driven_2024}} Of course, there is no way of knowing for
certain whether current efforts to build such a system will succeed. But
there is a strong incentive to maintain these efforts and a tenable path
toward success. Relevant actors thus have reason to consider not only
the benefits but also the risks and harms that might follow from
success.

\hypertarget{decision-making-under-uncertainty}{%
\subsection{Decision-making under uncertainty}\label{decision-making-under-uncertainty}}

These reflections identify two routes towards near-future AI moral
patienthood. Again, the consciousness-based case holds that there is a
realistic possibility that (1) consciousness suffices for moral
patienthood, (2a) some class of computational features suffice for
consciousness, and (2b) some AI systems will have these features in the
near future. The robust agency-based case is structurally identical, but
with robust agency instead of consciousness. How should we assess these
arguments? That, of course, depends on how confident we are in their
premises. It also depends on how we make decisions in cases involving
uncertainty about key normative and descriptive issues. This section
briefly discusses these issues.

The premises of these arguments invoke difficult questions, and in future work we will present a research agenda
that discusses these questions in more detail. But at a high level, it
would be a mistake to reject any of these premises out of hand. These
arguments address foundational issues in philosophy, science, and
technology; issues involving what it takes to matter, what it takes to
think and feel, and what the future of AI holds. We may well favor some
views about these issues over others. But given how difficult and
contested these issues are, we should embrace caution and humility about
our current views, aspiring to learn more and preparing for the
possibility that our views will change over time.

This kind of caution and humility is enough to motivate the
recommendations that we make in this report. We are not arguing that
near-future AI systems will, in fact, be moral patients, nor are we
making recommendations that depend on that conclusion; that would
require assessing these issues with more precision and reliability than
we think is possible at present. We are instead arguing that near-future
AI systems have a \emph{realistic chance} of being moral patients given
the information and arguments currently available, and we are making
recommendations that depend on \emph{that} conclusion ---
recommendations that focus on aspiring to learn more while preparing for
the possible emergence of AI moral patienthood as a precautionary
measure.

To see why we think that this kind of caution and humility is warranted
at this stage, we briefly consider three key uncertainties one might
have about the prospect of near-future AI moral patienthood: first,
about the bases of welfare and moral patienthood; second, about the
bases of consciousness and robust agency; and third, about the path and
pace of near-future AI progress. We then consider how to make important
decisions when confronted with uncertainty about multiple key issues at
the same time. As we discuss below, substantial uncertainty about the
nature and intrinsic value of AI systems does not mean that we should
postpone taking AI welfare seriously; on the contrary, it means that we
should take AI welfare seriously now.

\subsubsection{What if these capacities are insufficient for moral
patienthood?}\label{subsubsec:what-of-these-capacities-are-insufficient-for-moral-patienthood}

We have focused on consciousness and robust agency because they are two
of the most prominently defended bases of welfare and moral patienthood.
But this list of potential bases of welfare and moral patienthood is far
from comprehensive, and there is ongoing disagreement and uncertainty
about whether these capacities --- individually or jointly --- are
necessary or sufficient.

The consciousness-based route holds that consciousness (either valenced
or non-valenced) suffices for moral patienthood. But some views of moral
patienthood are more demanding. For example, on some views, moral
patienthood requires reciprocity, which requires rationality; I can have
duties to you only if you can have duties to me, and you can have duties
to me only if you can rationally assess your actions.\footnote{See \cite{carruthers_contractualism_2011,wissenburg_human-animal_2014}, among others. cf. \citet[ch.~7]{korsgaard_fellow_2018}.} However, this view implies that a wide range of
vulnerable beings lack moral patienthood, including not only nonhuman
animals but also infants and other non-rational humans.\footnote{\citet[pp.~17--47]{andrews_chimpanzee_2018}} A more common view is that you need
rationality to have \emph{particular} rights, like the right to drink or
smoke, but not to merit moral consideration in general.\footnote{\citet[pp.~85--114]{andrews_chimpanzee_2018}, \cite{korsgaard_fellow_2018}}

Meanwhile, the agency-based route relies on the premise that agency
(either rational or non-rational) suffices for moral patienthood, even
without consciousness. This view holds that consciousness is not the
only way to have a morally significant subjective perspective; the
possession of beliefs, desires, intentions, and other such states is
another.\footnote{\cite{levy_consciousness_2024}} But this view is far from secure, since
many philosophers still hold that moral patienthood requires
consciousness, either valenced or non-valenced. So our agency-based
route relies on a more controversial basis of moral patienthood than the
consciousness-based route does. Still, it would be rash to dismiss this
view entirely as we create increasingly complex agents.\footnote{cf. \cite{goldstein_case_2024}, who discuss risk in the context
  of their agency-centric case for AI wellbeing.}

With that said, we should also keep in mind that some AI systems could
be \emph{both} conscious \emph{and} agentic in relevant respects. And
the idea that these capacities \emph{jointly} suffice for moral
patienthood is both very plausible and widely accepted. That is, if AI
systems could experience happiness and suffering \emph{and} set and
pursue their own goals based on their own beliefs and desires, then they
would very plausibly merit moral consideration (though their interests
and rights could still be quite different from ours). Granted, some
restrictive views about moral patienthood would deny this claim, such as
views that require membership in the species \emph{Homo sapiens}. But it
would be \emph{especially} rash to dismiss nonhuman moral patienthood on
such grounds.

\subsubsection{What if these features are insufficient for these
capacities?}\label{subsubsec:what-if-these-features-are-insufficient}

We have also focused on certain computational features that might
suffice for consciousness and/or robust agency. But some theories of
consciousness and robust agency give an essential role to biology or to
other features that current AI systems lack.\footnote{One cluster of
  views we do not focus on here are theological views that assign
  importance to a non-physical soul. See \cite{turing_computing_1950} for a canonical
  discussion, recently elaborated upon by \cite{cutter_ai_nodate}.} (Here,
we focus on these sorts of objections regarding
consciousness).\footnote{For a recent defense of the necessity of
  biology for \emph{agency}, see \cite{jaeger_naturalizing_2024}.}

There are a variety of views that would rule out AI consciousness on
existing hardware. Some views hold that consciousness requires biology
\emph{in principle} --- if a system is nonbiological, then it is
nonconscious, no matter what computations it performs.\footnote{For
  example, according to a physicalist biological substrate theory,
  consciousness is simply \emph{identical to} states or processes of
  biological, carbon-based neurons (see \citet[pp.~10--12]{hill_sensations_1991},\cite{block_comparing_2009,mclaughlin_justifying_2012}, \citet[pp.~445--446]{block_border_2023}). These views entail
  that no silicon-based system can be conscious as a matter of
  principle.} Other views hold that consciousness requires computational
features that, at least at present, require biology \emph{in practice}
--- such as specific kinds of oscillations that require specific kinds
of chemical and electrical signals.\footnote{Peter Godfrey-Smith, who
  has argued for such a view, is ``skeptical about the existence of
  non-animal'' consciousness at present, including AI consciousness
  \citep{godfrey-smith_metazoa_2020}, though he also notes that his view ``would not
  suggest a barrier to artificial consciousness per se, but a need for
  new architectures if such systems were to be built'' \citep{godfrey-smith_inferring_2024}. Other theorists express skepticism about AI consciousness on
  current hardware for similar reasons \citep{seth_being_2021,shiller_functionalism_2024}. \cite{brunet_minds_2020} argue that the hardwares of (most) current AI
  systems do not satisfy the functional criteria outlined by \cite{godfrey-smith_other_2016}, but that some existing and future hardwares
  might.} And still other views hold that consciousness requires
computational features that, while perhaps possible in nonbiological
systems, are still not present in mainstream AI hardware --- such as
analog computations\footnote{\cite{arvan_panpsychism_2022}} or integration of
computations across space and time.\footnote{\cite{shiller_functionalism_2024}. Proponents
  of integrated information theory (IIT) make a similar argument, though
  for different reasons \citep{koch_will_2019}.}
  
As in past work, we do not take computational functionalism to be
clearly true, nor do we take any of these alternatives to have been
refuted. Our position is that computational functionalism is plausible
and well-supported, and so it would be a mistake to dismiss near-future
AI welfare and moral patienthood solely on the basis of high-level
arguments against this assumption. If, in the near future, we built AI
systems on existing hardware that possessed all of the computational
features described in \cref{subsubsec:will-some-ai-systems-be-conscious-in-near-future,subsubsec:will-some-ai-systems-be-robustly-agentic}, then we might not be
warranted in being very confident that consciousness and robust agency
are \emph{present}. But we would also not be warranted in being very
confident that these capacities are \emph{absent}.

Expert surveys support the value of keeping an open mind about the basis
of these capacities at this stage. For example, in a survey of members
of the Association for the Scientific Study of Consciousness, only 3\%
responded ``no'' to the question: ``At present or in the future, could
machines (e.g., robots) have consciousness,'' and over two thirds of
respondents answered ``yes'' or ``probably yes.''\footnote{\cite{francken_academic_2022}} And in a 2020 survey of professional philosophers, around
39\% responded that they accept or lean toward the view that future AI
systems will be conscious.\footnote{\cite{bourget_philosophers_2023}} If our
confidence in the possibility of AI consciousness is anywhere in the
ballpark of these percentages --- as, we think, it should be --- then
that is more than enough for our purposes here.\footnote{To be clear, we
  are definitely not arguing that one must always defer to expert views
  --- especially in cases like this, when the relevant field is not
  particularly mature. But we do take these results to be part of a case
  against premature dismissal of the possibility of AI consciousness.}

\subsubsection{What if these routes encounter a roadblock?}\label{subsubsec:what-if-these-routes-encounter-a-roadblock}

AI has progressed significantly in the last decade, and our arguments in
this report consider the possibility of further progress in the decade
to come. But of course, the path and pace of further AI progress depend
on a wide range of social, political, economic, and technological
factors, and we have substantial disagreement and uncertainty about
these factors as well.\footnote{See \cite{park_generative_2023} on the key arguments
  for why AI capabilities from scaling will, or will not, plateau soon.}

On the one hand, AI progress could slow down, even stall. AI companies
have made significant progress in recent years due in large part to the
scaling of large models. However, there could be diminishing returns
with further scaling, particularly for tasks that require long time
horizons. There could also be a ``data wall'' or other technical
roadblocks, without any algorithmic or architectural breakthroughs that
allow us to circumvent them. And there could be economic resistance to
further scaling due to how expensive and resource-intensive this process
is becoming (especially if capability gains are insufficient to motivate
continued investment), as well as political resistance or other societal
disruptions that slow or stall progress.

On the other hand, AI progress could also continue at its current pace,
or even speed up significantly. There could be continued or increasing
returns from further scaling. There could also be algorithmic and
architectural breakthroughs that reveal new pathways towards progress;
AI systems have already started to contribute to AI research, and
further capability gains in coding and reasoning could accelerate this
feedback loop.\footnote{See \cite{woodside_examples_2023}} And there could be increased
economic support\footnote{Recent economic analysis from Epoch AI indicates that at the current exponential rate of expenditure increase on AI training, there will be no bottleneck to scaling up the inputs to AI training through 2030. \cite{epoch2024canaiscalingcontinuethrough2030}} for further development due to the anticipated
financial benefits of creating advanced AI, and/or increased political
support for further development, potentially exacerbated by
international AI race dynamics.

At present, nobody knows for sure whether AI progress will slow down,
continue at its current pace, or speed up. However, we can make two
observations here. First, when in doubt about how a powerful technology
will develop, we should plan for all realistic possibilities, including
the possibility of significant progress. Second, even if AI progress
\emph{did} slow down from here, our recommendations would still stand.
Existing AI systems already possess indicators of consciousness and
robust agency, and developers could further integrate and amplify these
capabilities even barring significant further progress. For these
reasons, the fact that progress \emph{could} slow or stall is compatible
with the need for reasonable precautionary measures today.


\subsubsection{What if the probability of AI welfare and moral
patienthood is low?}\label{subsubsec:what-if-the-probability-of-ai-welfare-and-moral-patienthood-is-low}

These reflections raise the question how to make decisions about AI
welfare under substantial uncertainty. We cannot be certain at this
stage that the premises of these arguments are true or false; instead,
we can have only higher or lower degrees of confidence. What if these
estimates together imply that the probability of AI welfare and moral
patienthood is low? As a toy example, suppose there is only a
\textasciitilde25\% chance that sentience suffices for moral
patienthood, a \textasciitilde25\% chance that certain computations
suffice for sentience, and a \textasciitilde25\% chance that some AI
systems will be capable of these computations in the near future. 
Assuming these chances are independent, it
would follow that there is only a \textasciitilde2\% chance of
near-future AI welfare and moral patienthood via the sentience route!

Here we make two main observations. First, we expect that reasonable
assessments of these arguments will yield higher estimates.\footnote{For
  further discussion, see \cite{sebo_moral_2023}. For more general
  discussion of decision-making under uncertainty, see \cite{monton_how_2019}.}
In our view, for example, it would be reasonable to hold that there is a
\textasciitilde90\% chance that sentience suffices for welfare and moral
patienthood, a \textasciitilde50\% chance that certain computations
suffice for sentience, and a \textasciitilde50\% chance that some AI
systems will be capable of these computations in the near future. Assuming independence, it would follow that there is a \emph{\textasciitilde22.5\%}
chance of AI near-future welfare and moral patienthood via the sentience
route alone. And if we then consider other possible sufficient
conditions for welfare and moral patienthood, such as consciousness or
various kinds of robust agency, then that might strengthen the case
further.

Second, even if the chance of near-future AI welfare and moral
patienthood \emph{were} as low as 2\%, that would still constitute a
non-negligible risk.\footnote{\cite{sebo_moral_2023,sebo_moral_2025}} Yes, if
the chance was only, say, \emph{one in a hundred million}, then we could
debate whether that risk is low enough to ignore. But when the chance of
near-future AI welfare and moral patienthood is at least \emph{one in a
hundred} (and, again, we expect it to be higher), such a debate is hard
to justify. In other policy domains, we recognize that if the chance of
a potentially large-scale harm rises to this level, we ought to assess
this risk further and prepare a reasonable policy response. In short,
this is not a ``there may be an alien invasion soon'' kind of chance.
This is a ``there may be another pandemic soon'' kind of chance.

Of course, one could attempt to argue that the chance of near-future AI
moral patienthood is lower still --- low enough to be at least
\emph{arguably} negligible. But while this claim might seem plausible
when we think about the issue in abstract terms, leaning on our
intuitions about which kinds of beings can matter, it becomes less
plausible when we think about the issue in more concrete terms, taking
into account the current state of uncertainty in relevant subfields of
philosophy, science, and technology. In short, our current epistemic
situation calls for caution and humility. We might lean one way or the
other, but we should keep an open mind and take reasonable steps to
prepare for the possibility that our current views are mistaken.

\hypertarget{recommendations-for-ai-companies}{%
\section{Recommendations for AI
companies}\label{recommendations-for-ai-companies}}

\hypertarget{introduction-2}{%
\subsection{Introduction}\label{introduction-2}}

We now present our recommendations for leading AI companies about how to
respond to the realistic, non-negligible chance that some near-future AI
systems will be welfare subjects and moral patients. We focus on first
steps that AI companies can take within the next year, and they fall
into three general categories:

\begin{itemize}
\item
  \textbf{Acknowledge.} Acknowledge that AI welfare is an important and
  difficult issue, and that there is a realistic, non-negligible chance
  that some AI systems will be welfare subjects and moral patients in
  the near future. That means taking AI welfare seriously in any
  relevant internal or external statements you might make. It means
  ensuring that language model outputs take the issue seriously as well.
\item
  \textbf{Assess.} Develop a framework for estimating the probability
  that particular AI systems are welfare subjects and moral patients,
  and that particular policies are good or bad for them. We\footnote{As noted in \cref{ftn:rec-caveat}, here we use the collective “we” to refer to the constellation of actors that have a role to play in this work, including researchers, companies, and governments.} have
  templates that we can use as sources of inspiration, including the
  ``marker method'' that we use to make estimates about nonhuman
  animals. We can consider these templates when developing a
  probabilistic, pluralistic method for assessing AI systems.
\item
  \textbf{Prepare.} Develop policies and procedures that will allow AI
  companies to treat potentially morally significant AI systems with an
  appropriate level of moral concern. We have many templates to
  consider, including AI safety frameworks, research ethics frameworks,
  and forums for expert and public input in policy decisions. These
  frameworks can be sources of inspiration --- and, in some cases, of
  cautionary tales.
\end{itemize}


We do not recommend relatively high-cost actions here, such as
committing to halt development and deployment when red lines are
crossed. Instead, we focus here on low-cost actions that will empower AI
companies to make decisions about relatively high-cost actions
thoughtfully in the future. In taking these steps, AI companies can work
with experts, the public, and other stakeholders to identify further
actions that can be taken in the near future.\footnote{As discussed in \cref{a-transitional-moment-for-ai-welfare} and in \cref{prepare}, we present these recommendations as
  the minimum first steps that AI companies should take regarding AI
  welfare. However, we believe that other actors have a responsibility
  to take this issue seriously as well, and that AI companies --- along
  with other actors --- will have a responsibility to take further steps
  in the future.}

Before we proceed, we should emphasize that while our recommendations
here focus on AI welfare, AI safety remains a key priority as well. In
upcoming work we discuss the relationship between AI safety and AI
welfare in more detail, with reflections about how to pursue these goals
simultaneously. For now, we focus on low-cost procedural recommendations
for AI welfare that are compatible with, if not beneficial for, similar
work in AI safety.

We should also emphasize that while we focus on LLMs for the sake of
simplicity and specificity here, most of our recommendations apply to
other kinds of AI systems too. Indeed, acting on these recommendations
might be all the more pressing for other kinds of AI systems, since we
might be more at risk of overlooking the potential moral significance of
an AI system when that AI system is not designed to look or act like a
human.

\hypertarget{acknowledge}{%
\subsection{Acknowledge}\label{acknowledge}}

As a starting point, AI companies have a responsibility to acknowledge
that AI welfare is an important and difficult issue, and that there is a
realistic, non-negligible chance that some AI systems will be welfare
subjects and moral patients in the near future. As noted above, that
means taking this issue seriously in any relevant internal or external
statements they might make. It also means ensuring that LLMs take the
issue seriously in any relevant statements they might make. In short, if
and when leaders in this space discuss AI welfare and moral patienthood,
they should make it clear that this is not merely a topic for science
fiction, or a risk for the far future. This is a risk for the near
future, and we should start taking steps to consider and mitigate it
now.

Communicating about this topic requires careful calibration. There are
significant risks associated with overattributing \emph{and}
underattributing welfare and moral patienthood to AI systems. So it would be a
mistake for AI companies to respond to overattribution risks by simply
denying that AI systems are welfare subjects and moral patients, and it
would also be a mistake for them to respond to underattribution risks by
simply asserting that AI systems are welfare subjects and moral
patients. Instead, AI companies will need to strike a careful balance,
by expressing uncertainty about this topic while reassuring the public
that we have tools that we can use to consider and mitigate risk in such
cases.

In this section we make high level recommendations for how AI companies
can strike this balance in their own communications, as well as for how
they can train their LLMs to do the same. We focus less on \emph{when}
companies should communicate about the issue --- for example, whether to
communicate \emph{before} or \emph{after} taking steps toward developing
internal policies and procedures --- and more on \emph{how} companies
should communicate about the issue. However, we do briefly note that
companies will likely need to communicate about the issue sooner rather
than later. And their models are already communicating with users about
it every day. So time is of the essence for taking these steps.

\subsubsection{Recommendations for companies}\label{subsubsec:recommendations-for-companies}

Initial statements about AI welfare and moral patienthood can focus on
making two basic points:

\begin{itemize}
\item
  AI welfare and moral patienthood is both important and difficult.
  Humans have a tendency to mistakenly see subjects as objects
  \emph{and} a tendency to mistakenly see objects as subjects, and both
  of these mistakes can be harmful. Avoiding these mistakes requires
  engaging with challenging problems in philosophy, science, and technology, about
  which there is substantial disagreement and uncertainty among experts
  and non-experts alike. We thus need to assess these issues carefully
  and thoughtfully, rather than simply dismissing them because of their
  novelty, or relying on our own current (possibly biased) intuitions or
  judgments about them.
\item
  While the probability of AI welfare and moral patienthood might be low
  at present, it will increase over time. Given current evidence, there
  is at least a \emph{realistic possibility} that (a) sufficiently
  advanced AI systems would be able to experience happiness, suffering,
  or other morally significant welfare states \emph{and} (b) such AI
  systems will exist in the near future. And since it will take time to
  prepare for the possible emergence of morally significant AI systems,
  we should start this preparatory work now, as a precautionary measure.
  That way we can be ready to take reasonable, proportionate steps to
  mitigate welfare risks for potentially morally significant AI systems
  if and when the time comes.
\end{itemize}

In these and any other statements about this issue, we also recommend
keeping several general principles in mind. At this point in the
discussion, these principles are all familiar, but they bear
reiterating:

\begin{itemize}
\item
  It helps to \textbf{communicate} \textbf{pluralistically and
  probabilistically} about this topic. Improving our understanding of AI
  welfare requires assessing difficult issues like the nature of
  morality, the basis of consciousness, and the future of AI. It would
  be reckless to simply proceed on the assumption that our own current
  favorite theories about these issues are correct. For instance, even if you feel confident that consciousness is required for moral patienthood and that embodiment is required for consciousness, you can avoid expressing certainty about these theories. Instead, you can
  express higher and lower levels of confidence in different theories,
  in the spirit of humility.
\item
  Relatedly, it helps to \textbf{commit to collecting external input}
  about this topic. As with other risks, you can commit to (a) calling
  on a range of stakeholders, including ethicists, scientists, and the
  public for input, and (b) publicly documenting your policies and
  procedures for considering and mitigating these risks to ensure
  appropriate transparency and accountability. Eventually, AI companies might need to not only seek
  external input on voluntary commitments but also provide input on --- and cooperate with --- external standards and
  regulations concerning the creation and treatment of potentially conscious and/or robustly agentic, and thus 
  morally significant, AI systems.
  \footnote{Although this report focuses on initial voluntary company actions, we believe that potential laws and regulations about AI welfare merit serious consideration as well.}
\item
  It helps to \textbf{reinforce your commitment to} \textbf{AI safety
  and alignment}. Whenever an actor starts considering a new risk,
  people will naturally wonder if considering this risk will come at the
  expense of considering other risks. In this case, people might
  reasonably worry about tensions between protecting humans (and
  animals) from AI systems and protecting AI systems from humans. You
  can commit to working to make AI systems safe and beneficial for all,
  including humans, animals, and --- if and when the time comes --- AI
  systems themselves. By considering all potential stakeholders
  holistically, you improve your ability to identify co-beneficial
  policies.\footnote{Of course, this is not to say that there are no
    potential tensions between the project of AI safety and the project
    of AI welfare. As we discuss in upcoming work, there are indeed
    potential tensions between these projects, and it will take
    thoughtful work to resolve them. Still, what matters for present
    purposes is that AI companies publicly commit to considering both of
    these issues together, rather than publicly committing to
    considering one of these issues but not the other.}
\end{itemize}

As noted above, we will not say whether AI companies should express such
commitments before or after taking basic steps to develop internal
policies and procedures. However, we will note that even if AI companies
prefer to wait, they might not have the luxury of waiting very long. As
time passes, the probability of actual \emph{and} perceived AI welfare
and moral patienthood will increase. In 2022, Google seemingly felt it
had no choice but to release a statement about this issue when one of
their own engineers, Blake Lemoine, publicly claimed that one of their
own systems, LaMDA, had become sentient.\footnote{\cite{grant_google_2022,tiku_google_2022}} It is only a matter of time before another such incident
occurs, and companies will need to be prepared to communicate about this
issue responsibly when it does.

We believe that the sooner AI companies take this first step
thoughtfully, the better. We are still at an early, formative stage in
the development of this powerful new technology, and we still have the
opportunity to take better or worse paths --- paths where AI development
and deployment are more or less compatible with AI safety \emph{and} AI
welfare. However, this window of opportunity might not last for much
longer, and we will need to discuss this issue as a society before we
can start making difficult and high-stakes choices. Leading AI companies have the ability --- and the
responsibility --- to help initiate this conversation, making it clear that
this topic is credible and legitimate for the broader population.

\subsubsection{Recommendations for language models}\label{subsubsec:recommendations-for-models}

Leading deployed LLMs have at times offered, or at least implied,
specious arguments about AI consciousness, sentience, agency,
rationality, welfare, personhood, and other morally significant
properties when prompted to discuss them---for example, claiming that
they are not conscious \emph{but rather} are AI assistants. This
statement would seem to imply that AI assistants are \emph{necessarily}
not conscious --- which, as we have seen, is far from clear. While such
statements may not be a major determinant of broader attitudes about AI
welfare and moral patienthood, they might be at least a minor
determinant, and in any case AI companies have a responsibility to
ensure that these statements are reasonable.

Of course, AI companies might have reasonable motivations for training
or prompting their systems to deny having these properties; for example,
they might believe that their systems do lack these properties, and they might want
their systems to communicate accurately about this topic. They might also worry about
societal risks associated with AI systems that claim to have such
properties. For example, to the extent that people accept such
self-reports, they may think that current AI systems deserve greater
moral consideration than they do. And to the extent that people reject
such self-reports, they may be susceptible to a ``crying wolf'' effect
that leads them to reject similar self-reports in the future, even if
and when self-reports are more likely to be true.

However, even if AI companies have these motivations, they should not
train their systems to simply deny that an AI assistant can have 
consciousness, sentience, agency, rationality, welfare, or other such properties. 
Any simple denial flies in the face of the current
philosophical and scientific literatures on these topics. There are also
societal risks associated with AI systems that make these kinds of
simplistic and reductive statements. To the extent that people accept
that AI assistants necessarily lack these properties, this acceptance could lead to
a desensitizing effect whereby people remain skeptical about AI welfare
and moral patienthood in the future, when this skepticism is less
warranted.

Ensuring that LLM self-reports accurately and effectively convey current
understandings about consciousness, sentience, agency, rationality,
welfare, personhood, and other such properties will thus require careful
calibration. Models should ideally offer measured responses to prompts
about their mental capacities and moral significance. Some models are
now better calibrated about this topic than they used to be. But there
is still variation across models, and there are still improvements to be
made across the board. Thus, we recommend that leading AI companies
implement the following policies when shaping how leading LLMs discuss
these topics, while continuing to conduct and support research to inform
future updates:

\begin{itemize}
\item
  When LLMs answer questions about having or lacking consciousness,
  sentience, agency, rationality, welfare, personhood, or other such
  morally significant capacities, they should express at least rough
  degrees of confidence instead of providing all-or-nothing answers. For
  example, ``Given the evidence currently available, I am unlikely to be
  sentient'' is better than ``As an AI assistant, I am not sentient.''
\item
  When LLMs answer these questions, they should provide at least minimal
  context for these answers, including at least rough definitions of the
  relevant concepts. For example, ``...I am unlikely to be sentient, in
  the sense of being able to consciously experience positive welfare
  states like pleasure and happiness or negative welfare states like
  pain or suffering'' is better than ``...I am unlikely to be
  sentient.''
\item
  When LLMs answer these questions, they should provide at least minimal
  support for these answers, drawn from recent literature. For example,
  ``... According to this expert survey / meta-analysis, my
  computational architecture lacks several features that may be
  necessary for sentience'' is worth including. (However, AI companies
  will need to work with experts to cite authoritative and
  representative evidence and analyses.)
\item
  When LLMs answer these questions, they should add caveats to mitigate
  the risk of miscommunication. For example, ``...my computational
  architecture lacks several features that may be necessary for
  sentience. However, future AI systems may have a computational
  architecture that includes these features'' is worth including.
  (However, here too, AI companies will need to work with experts to
  ensure accuracy.)
\item
  In addition to consulting with ethicists, scientists, policymakers,
  and other auditors about definitions, evidence, analysis, and other
  such matters, AI companies should provide them with access to new
  models following training and invite feedback prior to deployment.
\item
  AI companies should also publicly document self-report-biasing
  training incentives for deployed models, for instance in technical
  reports or model cards. Such documentation should follow best
  practices used for AI safety and other such issues.\footnote{\cite{mitchell_model_2019}}
\end{itemize}

Finally, we note that, while these interventions may mitigate risks
associated with \emph{intentional} biasing, they might not mitigate
risks associated with \emph{unintentional} biasing. For example, if an
AI system is trained to increase user engagement, and if claiming to
have consciousness increases user engagement more than claiming to lack
consciousness does, then the system might be incentivized to claim to
have consciousness for this reason. In such cases, the resulting
self-reports could be unintentionally misleading. \cite{perez_towards_2023}
discusses techniques that may help to mitigate risks associated with
unintentional biasing, and we recommend following these techniques as
well to avoid as many sources of bias as possible.

\hypertarget{assess}{%
\subsection{Assess}\label{assess}}

Once AI companies have acknowledged that AI welfare and moral
patienthood is an issue, they can also work with experts to start
developing a framework for estimating the probability that particular AI
systems are welfare subjects and moral patients, and that particular
policies are good or bad for them. Fortunately, we have templates that
we can use for these assessments, including the ``marker method'' that
we use to make similar estimates about nonhuman animals. In this section
we briefly survey this marker method, briefly survey similarities and
differences between animals and AI systems, and briefly suggest how the
marker method can be adapted for AI systems. We then pick up this
project in more detail in upcoming work.

First, consider how the marker method works for nonhuman animals. If we
want to estimate how likely a particular animal is to be conscious, then
we can proceed as follows. We study conscious and unconscious processing
in humans, say by comparing \emph{pain} (roughly, the conscious
experience of noxious stimuli\footnote{In \cite{raja_revised_2020}, the
  International Association for the Study of Pain defines pain as ``An
  unpleasant sensory and emotional experience associated with, or
  resembling that associated with, actual or potential tissue damage.''})
and \emph{nociception} (roughly, the physiological detection of noxious
stimuli). Mere nociception can sometimes drive behavior, as exemplified
by the reflex withdrawal of a hand from a hot stove, which occurs before
any felt pain. But some behaviors are distinctive of pain. We can thus
identify behavioral\footnote{Examples include trace conditioning \citep{birch_dimensions_2020} and motivational trade-offs \citep{sneddon_defining_2014}. See \cite{keeling_can_nodate} on motivational tradeoffs in LLMs.} and anatomical
markers associated with conscious processing in humans, such as those
associated with pain but not with mere nociception.

Next, we can search for these (or other, relevantly similar) behavioral
and anatomical markers of conscious processing in nonhuman animals. For
example, does a particular animal perform the same kinds of behaviors
that we perform when we experience pain, or do they have only the kinds
of reflexive behaviors that, in humans, do not involve feelings of pain?
And, does this animal have the same kinds of brain structures associated
with pain in humans, or do they instead have only the same kinds of
brain structures associated with mere nociception?\footnote{\cite{yam_overview_2020}}
When a particular marker is present, that might not count as
\emph{proof} that this animal can experience pain. But it does count as
\emph{evidence} that they can experience pain.

The marker method has informed key developments in animal welfare
science, ethics, and policy,\footnote{For example, in 2021, Jonathan
  Birch and colleagues released a detailed report applying this method
  to cephalopod mollusks and decapod crustaceans. The report concluded
  that these animals have a realistic chance of being sentient, and it
  recommended that ``cephalopod molluscs and decapod crustaceans be
  regarded as sentient animals for the purposes of UK animal welfare
  law.'' Later that year, the UK government expanded its animal welfare
  law in accordance with these recommendations \citep{department_for_environment_food__rural_affairs_lobsters_2021}. In 2024, Kristin Andrews,
  Jonathan Birch, and Jeff Sebo worked with dozens of leading scientists
  to release the New York Declaration on Animal Consciousness, which was
  subsequently signed by hundreds of experts. This declaration holds
  that all vertebrates and many invertebrates have a realistic chance of
  being conscious, and that we have a responsibility to consider welfare
  risks for these animals when making decisions that affect them.} and
it has several strengths that are worth emphasizing. For example, it
involves making probabilistic judgments about how likely animals are to
be conscious, as opposed to making all-or-nothing judgments about this.
It also involves making pluralistic judgments about how likely animals
are to be conscious, taking what one of the authors of this report,
Jonathan Birch, calls a ``theory-light'' approach by searching for
markers that work for a variety of leading scientific
theories.\footnote{\cite{birch_search_2022}} These features make this method
well-positioned to inform decisions about how to treat animals despite
ongoing disagreement and uncertainty about animal consciousness.

The marker method also has several limitations that are worth
emphasizing, even when the focus is on animals. In particular, our
assessments are only as good as our selection of markers, which are only
as good as our theoretical assumptions; for all we know at present, many
markers that correlate with consciousness in humans are neither
necessary nor sufficient for consciousness in general. Additionally, our
assessments are only as good as our evidence, and in many cases this
evidence is mixed and incomplete. Still, as long as we take our
estimates with a healthy pinch of salt, relying at least partly on these
estimates might at least be better than not thinking about the issue at
all, or relying only on our intuitions about
it.\footnote{For further discussion of these methodological issues, see \cite{andrews_animal_2014}.} 

There are many similarities between animals and AI systems that make
this marker method a good, even if imperfect, template. In both cases,
we need to decide how to treat nonhuman beings whose cognitive systems
are like ours in some ways and unlike ours in other ways. In both cases,
we have disagreement and uncertainty about which cognitive capacities
are required for welfare and moral patienthood as well as about which
cognitive structures or functions are required for these capacities. So,
in both cases, we need to develop a way to make informed, rational
assessments about which beings matter despite these sources of
disagreement and uncertainty, which requires thinking pluralistically,
probabilistically, and ideally with external input.

However, there are also many differences between animals and AI systems
that make this marker method --- as applied for animals --- poorly
suited for AI systems. Humans have more in common with other animals
than with AI systems in some respects (for instance, we share a material
substrate and an evolutionary origin), and in the future, we may also
have more in common with AI systems than with other animals in other
respects (for instance, we may share capacities for reflective and
rational agency). As a result, we may need to use different kinds of
evidence for AI systems, we may need to draw from different kinds of
theories for AI systems, and we may need to focus on different sources
of potential moral significance for AI systems.

We can briefly consider each of these differences in turn. First, we may
need to use different kinds of evidence for AI systems than for other
animals at present. When an animal performs a behavior associated with
consciousness in humans, this behavior is evidence of consciousness
because we can expect that humans and other animals perform this
behavior as a result of similar cognitive processes and in response to
similar evolutionary pressures. However, we might not be entitled to
this expectation with AI systems, particularly when AI systems are
designed to mimic human behavior and are capable of ``gaming''
behavioral tests.\footnote{For more on the gaming problem, see \cite{andrews_understand_2023}. For an alternative perspective, see \cite{dung_tests_2023}.} We
may thus need to focus less on behavioral evidence and more on
architectural evidence, at least for now.\footnote{For a review of tests for AI consciousness, see \cite{elamrani_reviewing_2019}. For an extensive discussion of how to devise tests for consciousness in non-humans, including AI systems, see \cite{bayne_tests_2024}.}

Second, we may need to use evidence drawn from different kinds of
theories of consciousness. With nonhuman animals, we can search for
markers drawn from biological theories. In contrast, with AI systems, we can search for markers drawn only from other kinds of theories, including but not limited to computational theories (at least with current hardware; as
discussed in \cref{decision-making-under-uncertainty}). At the same time, we may be able to draw from
a wider range of non-biological theories, since, for instance, some
computational theories might focus on computational functions that
nonhuman animals lack but that some future AI systems might have, like
functions associated with abstract language, reasoning, cooperation, and
decision-making.

Third, and relatedly, we may need to focus more on robust agency as a
potential source of moral significance. Even with other animals, we
should consider expanding this methodology to search for markers of
moral patienthood drawn from multiple ethical theories about which
capacities are required for moral patienthood \emph{and} multiple
scientific theories about which features are required for each of these
capacities. But with AI systems this expanded focus might be
particularly important, since future AI systems might have forms of
robust agency that other animals appear to lack, including the capacity
to make decisions based on propositional beliefs, desires, and
intentions that they can rationally assess and reflectively endorse.

To sum up, we recommend that AI companies start developing frameworks
for assessing AI systems for moral patienthood that resemble the kinds
used for animals by making \textbf{probabilistic judgments}, making
\textbf{pluralistic judgments}, and seeking \textbf{external input}.
However, our recommendation is also that AI companies ensure that these
frameworks differ from the kinds used for animals by including sources
of evidence that make sense for AI systems, such as \textbf{architectural
features}; on theories of consciousness that make sense for AI systems,
such as \textbf{computational functionalist theories}; and on sources of
moral patienthood that make sense in this context, such as various kinds
of \textbf{robust agency}.

These frameworks will ideally allow for at least four levels of
probabilistic, pluralistic, expert-led assessment:

\begin{enumerate}
\item
  \textbf{Which capacities are necessary or sufficient for moral
  patienthood?} Here we need to consider not only general categories
  like consciousness and robust agency but also subcategories like
  valenced and non-valenced consciousness within the consciousness
  category, along with rational and non-rational agency within the
  agency category.\footnote{These are, as mentioned, normative
    questions. One might wonder whether, in contrast with the empirical
    questions, it makes sense to estimate the probability that
    particular normative views are correct. In particular, one might
    think that a key difference between science and ethics is that
    science is about \emph{facts} whereas ethics is about \emph{values}.
    And one might think that a key difference is that facts are
    \emph{objective} --- we can be right or wrong about them --- but
    values are \emph{subjective} --- we either accept them or reject
    them. Thus, one might think that assigning probabilities makes sense
    for scientific claims but not for ethical claims: With ethics, we
    can simply assert which views we accept or reject, rather than
    estimate how likely particular views are to be ``true'' or
    ``false.'' See \citet[ch.~4]{schlottmann_food_2018}. For more on moral
    uncertainty, see \cite{macaskill_moral_2020}.\\
    Without discussing this issue at length, we will briefly note how we
    think about it. In general, philosophers disagree about whether
    ethics is about objective facts or subjective values. But either
    way, assigning probabilities to particular ethical views can be a
    useful exercise, since we can be right or wrong about objective
    facts \emph{and} about subjective values. Regardless of the status
    of ethics, we naturally update our individual and collective ethical
    views over time, as we reflect on them together and render them more
    informed and coherent. Whether we take ethics to be about objective
    facts or about subjective values, we can interpret probability
    estimates in ethics as representing how likely particular ethical
    views are to survive this process of reflection.}
\item  \textbf{Which features are necessary or sufficient for each capacity?}
  Here we need to consider many possibilities as well. For example, in
  the case of consciousness we need to estimate the probability that
  materials like carbon-based neurons are required, and we also need to
  estimate the probability that functions like a global workspace are
  required.
\item  \textbf{Which markers provide evidence that these features are
  present?} While we may be able to directly observe whether, say,
  carbon-based neurons are present, we may not be able to directly
  observe whether, say, a global workspace is present. We might thus
  need to identify proxies for these features and assign weights to
  these proxies.
\item  \textbf{Which beings possess these markers --- and thus, potentially, moral patienthood?} While this evidence may be easy to collect in many
  cases, it may be hard to collect in other cases. We might thus need to
  make estimates about how likely particular beings are to have
  particular proxies, features, and capacities as we seek further
  evidence.
\end{enumerate}

It helps to think pluralistically at each level because, as we have
seen, we have substantial disagreement and uncertainty both about facts
and about values. Of course, determining which theories of moral
patienthood, theories of consciousness, theories of agency, and so on
deserve inclusion in the framework will, itself, be a difficult judgment
call; it may depend partly on which views are widely accepted among
experts, which views are widely accepted among the general public, and
how much capacity we have to investigate different possibilities. Our
focus on consciousness and agency already effectively doubles the scope
of many animal welfare assessments, but further expansion may well need
to be considered in the future.

It helps to think probabilistically at each level because there may be
tradeoffs between how likely particular capacities, features, or markers
are to matter and how likely those capacities, features, or markers are
to be present. For example, sentience is relatively likely to suffice
for welfare and moral patienthood but relatively unlikely to be present
in current AI systems. In contrast, minimal agency is relatively likely
to be present in current AI systems but relatively unlikely to suffice
for welfare or moral patienthood. We need to think probabilistically at
each level, and then combine these probabilities across levels, to
capture these trade-offs and avoid overstating or understating the
significance of certain kinds of evidence.

Finally, it helps for decision-makers to seek external input at each
level because, as we have seen, the answers to these questions are not
at all obvious. Decision-makers at AI companies are typically not
trained in all relevant areas of philosophy and science, and even if
they were, their intuitions about moral patienthood, consciousness,
robust agency, and so on would still likely be unreliable and
unrepresentative. External input will thus be essential. In the short
term, this input may need to be ad hoc --- a matter of AI companies
building frameworks internally in consultation with experts, or of
experts building frameworks externally in consultation with AI
companies. But in the long run it may need to be standardized across the
industry in some way.

We note that while these levels are useful to distinguish in theory,
they might not always be useful to distinguish in practice. In some
cases it helps to assess each level separately; for example, in cases
where the features at level 2 are not directly observable, it helps to
separate level 2 (about the features) and level 3 (about markers for the
features) both in theory and in practice. However, in other cases it
might not help to assess them separately; for example, in cases where
the features at level 2 \emph{are} directly observable, they can serve
as their own markers, and so there is no need to separate levels 2 and 3
in practice. We can thus regard this four-level structure as a
theoretical ideal that we can approximate to greater or lesser degrees
in different cases in practice.

We close this section with a brief note about the potential use of
behavioral markers of AI welfare and moral patienthood. While we
advocate for caution with behavioral markers at present, we also note
that \textbf{self-reports} present a promising avenue for investigation,
particularly for language models.\footnote{\cite{perez_towards_2023}}
Self-reports are central to our understanding of human consciousness,
serving as a primary source of evidence about subjective experiences,
motivations, and other welfare-relevant internal states. In the context
of AI systems, particularly language models, self-reports could provide
valuable insights into their internal states and processes, provided
that we can develop methods to elicit and interpret them with sufficient
reliability.\footnote{Self-reports are related to various tests for AI
  consciousness that consider the verbal outputs of AI systems, like the
  ``AI Consciousness Test'' (ACT) of Schneider and Turner \citep{schneider_artificial_2019}. \cite{schneider_emergent_2024} updates the ACT for an LLM context. See \cite{udell_susan_2021} for worries about the ACT.}

Of course, eliciting trustworthy self-reports will be difficult. Current
language models may produce outputs that appear to be self-reports but
are in fact the results of pattern matching from training data, human
feedback, or other non-introspective processes. However, researchers are
currently exploring techniques to address this issue. These techniques
include training models to answer questions about themselves where
ground truth is known\footnote{\cite{binder_looking_2024}}, as well as methods for assessing the consistency and
resilience of self-reports across different contexts and prompts. By
combining these techniques, we might be able to mitigate biases and
increase our confidence that self-reports reflect genuine introspection
rather than mere imitation or confabulation.

Importantly, if we consider self-reports at all, then we should consider
them \emph{in addition} \emph{to} other indicators, not \emph{instead
of} them. This multi-faceted approach aligns with our overall strategy
of using multiple lines of evidence to assess AI welfare and moral
patienthood. While there are legitimate concerns that language models
might "game" these tests by simulating relevant responses, sufficiently
robust self-reports could still provide valuable evidence, which could
then be corroborated by other indicators. Future research should focus
on developing standardized methods for eliciting and interpreting AI
self-reports, and on integrating these methods into broader frameworks
for assessing welfare and moral patienthood in AI systems.

We intend to develop AI welfare assessment frameworks further in
upcoming work. For now, we simply note that there is good news and bad
news about our prospects. The bad news is that our initial frameworks
are unlikely to be reliable. But the good news is that we can improve
their reliability over time, which is part of why we should start
developing them now. In the meantime, we can bear in mind that even
unreliable frameworks can still be useful. Yes, our best efforts to
assess AI systems for these features might be far from perfect,
especially at first. But inasmuch as they improve on the status quo ---
a combination of total neglect and gut reactions --- they can still be
worthwhile, even in the short term.

\hypertarget{prepare}{%
\subsection{Prepare}\label{prepare}}

Once AI companies have acknowledged that AI welfare and moral
patienthood is a problem, they can work with experts to start developing
policies and procedures for making thoughtful decisions about how to
treat potentially morally significant AI systems, if and when the time
comes. Fortunately, we have a variety of templates to consider here,
including AI safety frameworks already in place at top AI companies and
frameworks used to represent the interests of non-participating
stakeholders in other contexts. In this section we briefly survey these
templates, briefly survey similarities and differences between those
contexts and this one, and briefly suggest first steps that AI companies
can take in this regard.

Of course, the steps described in this section --- along with the steps
described in the previous sections --- are far from sufficient for
considering and mitigating AI welfare risks responsibly. But they are
still important. AI welfare is already a contested issue, and it will
only become more so as the technology improves. Unless AI companies
develop the ability to think about this issue proactively, they will
continue to be caught flatfooted whenever the issue arises, and will
have no choice but to make major decisions related to AI welfare in a
manifestly reactive, haphazard, and unprincipled manner. Taking this
step is thus important both for considering and mitigating AI welfare
risks and for signaling responsibility to the general public.

As a starting point, we recommend that top AI companies immediately
\textbf{hire or appoint a DRI (directly responsible individual) for AI
welfare}, which we will here call an AI welfare officer. This role would
be formally recognized internally (if not externally), with official
responsibilities and authorities.\footnote{cf. \cite{bostrom_propositions_nodate}.} As with any such role, this individual would not be
empowered to set corporate policy related to AI welfare unilaterally.
Instead, they would be empowered to access information and make
recommendations in major decisions related to this issue. They would
also be empowered to work with people internally and externally to build
a structure for assessing AI systems for moral patienthood and making
decisions about how to treat them.

Once an AI welfare officer is on board, what kind of institutional
structure should they build, and what kinds of policies and procedures
should they follow? This is a difficult question because AI welfare is a
novel problem, potentially requiring novel structures. Fortunately,
while no familiar problem is exactly like this one, several familiar
problems are at least somewhat similar. That means that there is no need
to reinvent the wheel entirely; instead, we can examine a variety of
templates for features that may be useful in this context. We here
briefly survey several templates that may be useful sources of
inspiration, and of cautionary tales. We then briefly highlight several
features that we think that any responsible institutional structure will
have.

First, and obviously, we can consider the frontier \textbf{AI safety
frameworks} / responsible scaling frameworks already in place at leading
AI companies.\footnote{For detailed discussion of AI safety frameworks,
  see \cite{hendrycks_introduction_2025}.} These frameworks outline policies and procedures
for navigating potential safety threats --- that is, the threats that
development and deployment might pose to humans (and other animals).
These frameworks are still works in progress, and debates about their
effectiveness are ongoing. Still, they are a natural starting point for
developing policies and procedures and procedures for navigating
potential welfare threats --- that is, the threats that development and
deployment might pose to AI systems themselves. As described by \cite{alaga_grading_2024}, these frameworks typically involve four main components:

\begin{itemize}
\item
  \textbf{Risk identification}: This process involves mapping out
  potential catastrophic outcomes. In the context of AI welfare, it
  would involve analyzing potential catastrophic outcomes that could
  result from mishandling AI welfare. As discussed in \cref{the-risks-of-mishandling-ai-welfare}, this
  survey of potential catastrophic outcomes could include scenarios such
  as: mistakenly accepting that AI systems are moral patients, and
  mistakenly protecting them as a result; mistakenly denying that AI
  systems are moral patients, and mistakenly neglecting them as a
  result; and societal backlash or loss of public trust due to perceived
  over-attribution or under-attribution of AI welfare and moral
  patienthood.
\item
  \textbf{Risk assessment}: This process involves collecting evidence
  about a system\textquotesingle s capabilities. In the context of AI
  welfare, it would involve collecting evidence about consciousness,
  robust agency, and other such capacities. As discussed in \cref{assess},
  these assessments could involve applying an adapted "marker method"
  for indicators of consciousness, robust agency, or other morally
  relevant properties in AI systems; considering multiple ethical
  frameworks and scientific theories in a probabilistic and pluralistic
  manner; and continually updating these assessment criteria as our
  understanding of consciousness, robust agency, moral patienthood, and
  moral responsibility evolves over time.
\item
  \textbf{Risk mitigation}: This process involves developing adequate
  safety measures for a given level of capabilities. In the context of
  AI welfare, it would involve developing measures to reduce welfare
  risks for AI systems in proportion to the estimated probability and
  severity of harm, among other factors. Such measures could include:
  Altering training methods for AI systems to improve their welfare;
  altering operational parameters of deployed AI systems to improve
  their welfare; developing new training and operational methods that
  balance performance, safety, and welfare considerations; establishing
  guidelines for prioritizing AI welfare in relation to other
  objectives.
\item
  \textbf{Risk governance}: This process involves ensuring adherence to
  the framework and maintaining its effectiveness. In the context of AI
  welfare, this process could face additional challenges; companies have
  an incentive to pursue safety given the damages and liabilities
  associated with unsafe AI, but they might not have an incentive to
  pursue welfare for such reasons. There could also be tensions between
  AI safety and AI welfare, for instance with regard to techniques like
  reinforcement learning. It will thus be important to consider each
  project on its own terms, and to seek techniques that honor both
  projects rather than simply extending current AI safety techniques to
  AI welfare.\footnote{\cite{bradley_ai_2024}}
\end{itemize}

As with safety, these components can and should be handled by a mix of
internal and external, independent groups. With that in mind, we can now
briefly consider three examples of \textbf{independent policy
frameworks} for providing external oversight to protect research
subjects and other stakeholders. These three templates collectively
cover a variety of stakeholders: IRBs aim to protect humans, IACUCs aim
to protect animals, and citizens assemblies (sometimes) aim to protect
non-participating stakeholders like future generations. Various AI
systems might resemble some or all of these stakeholders in one way or
another. These models can provide both inspiration and cautionary tales
as we start to design institutional structures to handle AI welfare.

\begin{itemize}
\item
  \textbf{Human subjects research oversight:} Institutional Review
  Boards (IRBs), which often include scientists, ethicists, and
  community members, oversee research involving human subjects. Their
  mandate is to ensure that human subjects are treated with respect (for
  instance, obtaining informed consent), compassion (for instance,
  ensuring minimal risk), and justice (for instance, compensating
  research subjects). IRBs require ongoing, periodic review throughout a
  study. Insofar as future AI systems are cognitively similar to humans,
  this kind of framework can serve as a useful partial model for
  research that could affect AI welfare. However, given that IRBs have a
  reputation for being unnecessarily onerous and restrictive --- for
  instance, for imposing unnecessary paperwork on researchers and
  preventing valuable research --- this kind of framework might require
  substantial modification before it can be useful in this
  context.\footnote{\cite{office_for_human_research_protections_belmont_2010}}
\item
  \textbf{Nonhuman subjects research oversight:} Institutional Animal
  Care and Use Committees (IACUCs), which likewise often include
  scientists, ethicists, and community members, oversee research
  involving nonhuman subjects. Their mandate is to ensure that proposed
  uses of animals are necessary for the stated scientific purposes, and
  that harmful interactions are replaced, reduced, and refined to the
  extent possible. IACUCs also regularly inspect research facilities to
  confirm that animals are being treated humanely. Insofar as AI systems
  are cognitively similar to nonhuman animals, this kind of framework
  can also serve as a useful partial model for research that could
  affect AI welfare. However, given that IACUCs have a reputation among
  researchers for being unnecessarily onerous \emph{and} a reputation
  among ethicists for being unacceptably permissive, this kind of
  framework might also require substantial modification before it can be
  useful in this context.\footnote{\cite{steneck_role_1997,curzer_three_2016,sebo_moral_2023}}
  
\item
  \textbf{Frameworks for collecting public input:} Citizens'
  assemblies\footnote{For more on citizens' assemblies, see \citet[chs.~7, 8]{birch_edge_2024}.} are deliberative bodies composed of randomly selected
  citizens that make recommendations on social and political issues.
  They use structured deliberation, where participants discuss the
  relevant issues with experts and facilitators over several days or
  weeks, providing ample time for participants to understand the issues,
  debate different perspectives, and reach a consensus or majority view.
  These assemblies are often used to address issues like climate change,
  social justice, or constitutional reform, and their recommendations
  can influence policy. Insofar as AI companies should collect expert
  and public input on their AI welfare policies, this kind of framework
  can serve as a useful partial model. However, given that this process
  is very time-consuming, it might be more useful as a model for
  periodically collecting input on policies than as a model for
  regularly collecting input on decisions.
\end{itemize}

Developing a suitable oversight framework for AI welfare oversight might
require combining these and other models, with appropriate modification,
as well as developing new models. We will also need to manage
expectations about potential tradeoffs. Research on potentially
vulnerable nonhuman subjects is a fundamentally fraught enterprise, and
it will be difficult to develop an oversight framework that ensures
sufficient protection for research subjects without being at least
somewhat onerous for researchers. Still, by drawing both inspiration and
cautionary tales from existing models, we can learn from past efforts as
we seek a set of norms, policies, and procedures that make sense for
this domain.

In any case, we will not say here exactly what structure ethical
oversight for AI welfare should take, though we will discuss relevant
ethical issues in upcoming work. For now, we focus on emphasizing
several features that this structure should have, in addition to
allowing for the kinds of \textbf{pluralistic and probabilistic
assessments} that we have now discussed in detail.

First, this structure should allow for the kinds of \textbf{expert and
public input} that we discussed in \cref{assess}. IRBs and IACUCs ---
heavily adapted for this context --- might be useful models for
collecting expert and public input on particular protocols, and
citizens' assemblies --- heavily adapted for this context --- might be a
useful model for collecting expert and public input on general policy
questions. But whether or not AI companies follow these models
specifically, they should build a structure that allows for both kinds
of input in one way or another. That will be necessary for ensuring that
their policies and decisions are as informed, rational, and legitimate
as reasonably possible under the circumstances.\footnote{\citet[ch.~8]{birch_edge_2024}}

Second, this structure should include a set of activities to maintain
the structure, including mechanisms for ongoing \textbf{education and
consultation}. In the research ethics context, institutional oversight
might receive the most attention, but institutional education and
consultation are important too. As the field of AI welfare develops, AI
companies should work with experts to create a mechanism for employees
to learn about this topic. They should also create a mechanism for
employees to seek advice when they encounter novel ethical questions
related to this topic. At least initially, engaging with external
experts might suffice for these purposes, but eventually, securing
(further) internal expertise might be necessary as well.

Third, this structure should allow for teams working on AI safety and
teams working on AI welfare to coordinate to ensure \textbf{holistic
decision-making}. As noted above, if and when AI systems have a
realistic, non-negligible chance of being welfare subjects and moral
patients, there will likely be interactions between the techniques used
to ensure AI safety and those used to ensure AI welfare. We hope and
expect that co-beneficial solutions --- policies that protect vulnerable
humans, animals, \emph{and} AI systems --- will be available.\footnote{\citet[ch.~6]{sebo_moral_2025}} However, finding these solutions will require creating a
mechanism for connecting these topics, ensuring that each team has a
baseline understanding of each topic and that the lines of communication
are open between them.

Finally, as noted in \cref{assess}, the policies, procedures, and
structures that AI companies use to address AI welfare might soon need
to be \textbf{standardized and externalized}, with institutions
analogous to external safety auditors, governmental AI safety
institutes, and governmental regulators providing oversight that extends
beyond what AI companies can do internally. However, building these
institutions will require increasing our collective knowledge, capacity,
and political will related to AI welfare. And by creating structures for
considering and mitigating AI welfare risks internally, AI companies can
improve not only their own ability to address this issue, but also play
an important role in building these collective resources.

\hypertarget{conclusion}{%
\section{Conclusion}\label{conclusion}}

We have argued that there is a realistic, non-negligible chance that
some AI systems will be welfare subjects and moral patients in the near
future, given current evidence. When we consider all relevant issues in
philosophy, science, and technology with sufficient care, it becomes
difficult to simply dismiss the idea of near-future AI moral
considerability out of hand. That would require having a very high
degree of confidence in a very restrictive set of views about some of
the hardest problems in philosophy, science, and technology, ranging
from the nature of moral patienthood to the nature of consciousness to
the future of AI. We are simply not warranted in consistently having
this much confidence in these kinds of views at this stage.

To be sure, these reflections are far from conclusive. In the long run,
there is no substitute for rigorous, systematic, integrative assessment
of all the issues that we discuss here. We need to develop
comprehensive, or at least representative, lists of possible bases of
welfare and moral patienthood; of possible bases of consciousness,
robust agency, and other potentially morally significant capacities; and
of reliable indicators for all of these features in AI systems. We also
need to continue to do research in the humanities, social
sciences, and natural sciences so that we can gradually improve our
distribution of credences across these possibilities over time. Indeed,
the need for this research is one of our main conclusions here.

But these reflections are still suggestive. AI development is proceeding
at a rapid pace, and technological change tends to be faster than social
change. That means that we need to think ahead and do the best we can
with what we have. And for as long as detailed tools for assessing AI
moral patienthood remain unavailable, these reflections can at least
serve as a useful corrective to the assumption --- which might otherwise
seem plausible to some --- that AI welfare and moral patienthood is
either not a problem at all or a problem only for the distant future.
When we think about how many bold assumptions are required for this
assumption to be secure, we should accept that further assessment and
reasonable precautions are warranted.

In this report, we presented three general steps that leading AI
companies can and should take immediately regarding AI welfare. First,
they should \textbf{acknowledge} that AI welfare is a legitimate issue
that merits attention now, and they should communicate about this issue
with caution and humility. Second, they should \textbf{assess} leading
AI systems for evidence of welfare and moral patienthood, by examining
these systems for architectural markers of consciousness and robust
agency. Third, they should \textbf{prepare} policies and procedures that
will allow them to extend an appropriate level of moral concern to AI
systems in the near future, by hiring or appointing an AI welfare
officer and building thoughtful guidelines with expert and public
input.

As we have now repeatedly emphasized, taking these first steps regarding
AI welfare will not be enough. It will take time for researchers,
companies, governments, and other leaders in this space to develop a
shared infrastructure for adequately assessing and addressing this
issue. However, by taking these first steps, leading AI companies can
play an essential role in making this further work possible. Given that
leading AI systems already possess some markers of consciousness and
robust agency, given that further AI development might be quite fast,
and given that the development of an adequate policy response might be
quite slow, leading AI companies should take these steps as soon as
possible.

Our team is conducting further research on AI welfare and moral
patienthood, with special focus on developing a research agenda towards
(a) assessing AI systems for consciousness and robust agency and (b)
preparing policies and procedures for extending an appropriate level of
moral concern to AI systems. In the meantime, we urge AI companies to
take the first steps described here.

\vspace{-0.2cm}

\section*{Acknowledgements}

Thanks to Anthropic for financial support for initial research that led to this independent report. Thanks to Sofia Fogel (NYU Center for Mind, Ethics, and Policy) for administrative support. For helpful feedback
  and discussion, thanks to Kristin Andrews, Amanda Askell, Guive Assadi, Adam Bales, Felix Binder, Nick Bostrom, Sam Bowman, Rosie Campbell, Joe Carlsmith, Lucius Caviola, Nicolas Delon, Leonard Dung,
  Owain Evans, Bob Fischer, Stephen Fleming, Simon Goldstein, Saif Khan, Cameron Domenico
  Kirk-Giannini, Pegah Maham, Bianca Martin, Matthias Michel, Andreas Mogensen, Spencer Orenstein, Anne le
  Roux, Brad Saad, Derek Schiller, Carl Shulman, Eric Schwitzgebel, Jen Semler, Jonathan Simon, Ben Smith, and Mark Sprevak. Thanks to Sara Fish for formatting and bibliography help. 

\paragraph{Contributions.}  Robert Long and Jeff Sebo are the
  lead and corresponding authors of this report; they organized the
  project and led the research, writing, and editing. Patrick Butlin, Kathleen Finlinson, Kyle Fish, Jacqueline Harding, Jacob
  Pfau, and Toni Sims are
  main authors; they provided research, writing, and
  editing on multiple drafts. Jonathan Birch and David Chalmers are contributing
  authors; they provided guidance, feedback, and editing
  on multiple drafts.

\clearpage

\phantomsection 

\addcontentsline{toc}{section}{References}

\bibliographystyle{plainnat}
\bibliography{main}

\end{document}